\documentclass[aps,prd,groupedaddress,showpacs,showkeys]{revtex4}
\usepackage{amssymb}
\usepackage{amsmath}
\usepackage{amsfonts}
\usepackage{epsfig}
\usepackage{graphicx}
\usepackage{tabularx}

\newcommand{\beaa}{\begin{eqnarray*}}
\newcommand{\enaa}{\end{eqnarray*}}
\newcommand{\bea}{\begin{eqnarray}}
\newcommand{\ena}{\end{eqnarray}}
\newcommand{\seq}{\begin{subequations}}
\newcommand{\sen}{\end{subequations}}
\newcommand{\eq}{\begin{eqnarray}}
\newcommand{\en}{\end{eqnarray}}

\def\shiftdown#1{#1\llap{\lower.04ex\hbox{#1}}}

\newcommand{\ra}{\rangle}
\newcommand{\la}{\langle}

\def\arraystretch{1.5}

\begin{document}

\title{Dilaton in a soft-wall holographic approach to mesons and baryons}
\noindent
\author{
Thomas Gutsche$^1$,
Valery E. Lyubovitskij$^1$
\footnote{On leave of absence
from Department of Physics, Tomsk State University,
634050 Tomsk, Russia},
Ivan Schmidt$^2$,
Alfredo Vega$^2$
\vspace*{1.2\baselineskip}\\}

\affiliation{
$^1$ Institut f\"ur Theoretische Physik,
Universit\"at T\"ubingen, \\
Kepler Center for Astro and Particle Physics,
\\ Auf der Morgenstelle 14, D-72076 T\"ubingen, Germany
\vspace*{1.2\baselineskip} \\
\hspace*{-1cm}
$^2$ Departamento de F\'\i sica y Centro Cient\'\i
fico Tecnol\'ogico de Valpara\'\i so (CCTVal), Universidad T\'ecnica
Federico Santa Mar\'\i a, Casilla 110-V, Valpara\'\i so, Chile
\vspace*{1.2\baselineskip}\\}

\date{\today}

\begin{abstract}

We discuss a holographic soft-wall model developed for the
description of mesons and baryons with adjustable quantum numbers
$n, J, L, S$. This approach is based on an action which describes
hadrons with broken conformal invariance and which incorporates
confinement through the presence of a background dilaton field. 
We show that in the case of the bound-state problem 
(hadronic mass spectrum) two versions of the model with a positive and 
negative dilaton profile are equivalent to each other by a special 
transformation of the bulk field. We also comment on recent works 
which discuss the dilaton sign in the context of soft-wall approaches.

\end{abstract}



\maketitle

\section{Introduction}

Based on the correspondence of string theory in anti-de Sitter (AdS)
space and conformal field theory (CFT) in physical
space-time~\cite{Maldacena:1997re}, a class of AdS/QCD approaches
was recently successfully developed for describing the phenomenology
of hadronic properties. In order to break conformal invariance and
incorporate confinement in the infrared (IR) region two alternative
AdS/QCD backgrounds have been suggested in the literature: the
``hard-wall'' approach~\cite{Hard_wall1}-\cite{Hard_wall3}, based on
the introduction of an IR brane cutoff in the fifth dimension, and
the ``soft-wall'' approach~\cite{Soft_wall1}-\cite{SWplus}, based on
using a soft cutoff. This last procedure can be introduced in the
following ways: i) as a background field (dilaton) in the overall
exponential of the action (``dilaton'' soft-wall model),
ii) in the warping factor of the AdS metric (``metric'' soft-wall model),
iii) in the effective potential of the action. 
These methods could in principle be equivalent to each other due to 
a redefinition of the bulk field involving the dilaton or by a redefinition 
of the effective potential.
Moreover, in Ref.~\cite{No_wall} a so-called no-wall holographic model was
recently proposed. This approach is motivated by the soft-wall model,
containing the dilaton field in the exponential
prefactor~\cite{Soft_wall1,Soft_wall2aa,Soft_wall2ab,Soft_wall2ac,%
Soft_wall2b}. After a special
dilaton transformation, the prefactor is moved to the effective
potential, where the dilaton field is replaced by a dilaton condensate.
We would like to stress that the
equivalence between the different versions of the soft-wall models is important
to guarantee that they describe the same physics. In particular, below
we show how to match the ``metric'' soft-wall model to the ``dilaton'' 
soft-wall model that both are equivalent in describing the hadronic 
phenomenology. Application of the ``metric'' models to other problems, 
like e.g. the hierarchy problem in the Randall-Sundrum 
model~\cite{Randall:1999ee} goes beyond the scope of the present discussion. 

In the literature there exist detailed discussions of the sign
of the dilaton profile in the dilaton exponential
$\exp(\pm\varphi(z))$~\cite{Soft_wall1,Soft_wall2aa,%
Soft_wall2ab,Soft_wall2ac,Soft_wall2b,%
SWminus,SWplus} for the soft-wall model (for a discussion of the
sign of the dilaton in the warping factor of the AdS metric see
Refs.~\cite{Soft_wall3a}). The negative sign was suggested in
Ref.~\cite{Soft_wall1} and recently discussed in Ref.~\cite{SWplus}.
It leads to a Regge-like behavior of the meson spectrum, including a
straightforward extension to fields of higher spin~$J$. Also, in
Ref.~\cite{SWplus} it was shown that this choice of the dilaton sign
guarantees the absence of a spurious massless scalar mode in the
vector channel of the soft-wall model. We stress that alternative
versions of this model with positive sign are also possible,
but they are restricted in applications. In particular, the version
with a positive dilaton is obtained by redefinition of the bulk field $V(x,z)$
as $V(x,z) = \exp(\varphi) \tilde V(x,z)$, where the transformed field
corresponds to the dilaton with an opposite profile. It is clear
that the underlying action changes, and extra potential terms are
generated depending on the dilaton field. These potential terms
vanish only in the case of spin-1 modes. In Refs.~\cite{SWminus} the
sign of the dilaton was just changed without adding the
corresponding potential terms, which is true only for total spin
$J=1$ and not correct for higher spins. Here we agree with the
criticism of Ref.~\cite{SWplus}. 

In Refs.~\cite{Soft_wall2aa,Soft_wall2ab,Soft_wall2ac,%
Soft_wall2b,Soft_wall6,Soft_wall8,Soft_wall9}
two equivalent versions of the soft-wall model for the study of
the bound-state problem (hadronic mass spectrum) with a positive
$\exp(\varphi)$~\cite{Soft_wall2aa,Soft_wall2ab,Soft_wall2ac,%
Soft_wall2b,Soft_wall6} and a
negative $\exp(-\varphi)$~\cite{Soft_wall8,Soft_wall9} dilaton
exponential have been developed. These approaches are based on
slightly different actions, which have the advantage of possible
applications to mesons and baryons with adjustable quantum numbers
of radial excitation, orbital and total angular momentum.
In the study of the bound state problem
both versions use effective actions quadratic in the bulk fields, which
are equivalent due to a redefinition of the bulk field containing the dilaton.
As a result both approaches arrive at the same equation of motion (EOM)
for the mode dual to hadrons with specific quantum numbers. Therefore, when
performing the matching of matrix elements in the soft-wall model
and light-front QCD, a precise mapping of the bulk modes
in the AdS fifth dimension to the hadron light-front wave functions can
be performed (see details in
Refs.~\cite{Soft_wall2aa,Soft_wall2ab,Soft_wall2ac,Soft_wall2b,LFH}).
The main objective of this paper is to demonstrate how to
correctly derive the versions of the soft-wall model with different signs of the
dilaton profile for the study of the hadronic mass spectrum.
We also consider an extension of the soft-wall model to
mesons and baryons with higher total angular momenta $J$.

When considering dynamical hadron quantities encoding a
nontrivial momentum dependence (e.g. form factors, parton distributions) the
sign of the dilaton profile becomes sufficient to guarantee fulfillment of the
boundary conditions at $z \to \infty$. As it was shown in Ref.~\cite{SWplus}
the soft-wall model with a positive dilaton cannot be directly applied to the
calculation of the bulk-to-boundary propagators of AdS fields by solving the
corresponding differential equation in the $z$-dimension. The corresponding
propagators $V(q^2,z)$ are dual to the external QCD currents and depend on the
holographic coordinate $z$ and the transverse momentum squared $q^2$.
In particular, the soft-wall model with a negative dilaton profile gives a
bounded solution for the bulk-to-boundary propagator of the AdS vector field at
$z \to \infty$. It also supplies the correct normalization
$V(0,z)=1$ at $q^2=0$ in accordance
with gauge invariance. The version with a positive profile, however, gives a
divergent solution. This means that the positive version of the soft-wall model
in its original form cannot be applied for the calculation of
AdS bulk-to-boundary propagators and dynamical hadron properties.
Therefore, the version with a negative dilaton can be applied without any
additional modification (or restriction) to the calculation of both mass
spectrum and dynamical properties of hadrons and we certainly prefer this
realization of the soft-wall model.

\section{Bosonic case}

\subsection{Scalar field}

First we demonstrate how to correctly derive the versions of soft-wall model 
with different signs of the dilaton profile for the study of hadronic mass 
spectrum. Afterwards we consider the fermionic
field and the extension to higher values of the total angular momentum $J$.
We consider the propagation of a scalar field $S(x,z)$ in $d+1$
dimensional AdS space. The AdS metric is specified by
\eq
ds^2 = 
g_{MN} dx^M dx^N = \eta_{ab} \, e^{2A(z)} \, dx^a dx^b = e^{2A(z)}
\, (\eta_{\mu\nu} dx^\mu dx^\nu - dz^2)\,, \hspace*{1cm}
\eta_{\mu\nu} = {\rm diag}(1, -1, \ldots, -1) \,,
\en
where $M$ and
$N = 0, 1, \cdots , d$ are the space-time (base manifold) indices,
$a=(\mu,z)$ and $b=(\nu,z)$ are the local Lorentz (tangent) indices,
$g_{MN}$ and  $\eta_{ab}$ are curved and flat metric tensors, which
are related by the vielbein $\epsilon_M^a(z)= e^{A(z)} \,
\delta_M^a$ as $g_{MN} =\epsilon_M^a \epsilon_N^b \eta_{ab}$. Here
$z$ is the holographic coordinate, $R$ is the AdS radius, and $g =
|{\rm det} g_{MN}| = e^{2 A(z) (d+1)}$. In the following we restrict
ourselves to a conformal-invariant metric with $A(z) = \log(R/z)$.

The actions for the scalar field $(J=0)$ with
a positive or negative dilaton are~\cite{Soft_wall2aa,Soft_wall2b}
\eq
S^+_0 = \frac{1}{2} \int d^dx dz \, \sqrt{g} \, e^{\varphi(z)} \,
\biggl[ g^{MN} \partial_M S^+(x,z) \partial_N S^+(x,z)
- \mu_0^2 \, S^+(x,z)S^+(x,z) \biggr]
\en
and~\cite{Soft_wall8}
\eq
S^-_0 = \frac{1}{2} \int d^dx dz \sqrt{g} e^{-\varphi(z)}
\biggl[ g^{MN} \partial_M S^-(x,z) \partial_N S^-(x,z)
- \Big(\mu_0^2 + \Delta V_0(z)\Big) \, S^-(x,z)S^-(x,z) \biggr] \,.
\en
The superscripts $+$ and $-$ correspond to the cases of
positive and negative dilaton, respectively, and 
$\varphi(z) = \kappa^2 z^2$. 
The actions are equivalent to each other, which is obvious
after performing the bulk field redefinition:
\eq
S^\pm(x,z) = e^{\mp\varphi(z)} S^\mp(x,z) \,.
\en
The difference between the two actions is absorbed in the effective
potential $\Delta V_0(z) = e^{-2A(z)} \Delta U_0(z)$,
where
\eq
\Delta U_0(z) = \varphi^{\prime\prime}(z)
+ (d-1) \, \varphi^{\prime}(z) A^{\prime}(z) \,,
\en
with $F^{\prime}(z) = dF(z)/dz$,
$F^{\prime\prime}(z) = d^2F(z)/dz^2$ and $F=\varphi, \, A$.
The quantity $\mu_0^2 R^{2}= \Delta (\Delta - d)$ is
the bulk boson mass, where $\Delta$ is the dimension of the
interpolating operator dual to the scalar bulk field.
For the case of the bulk fields dual to the scalar mesons
$\Delta = 2 + L$, where $L = {\rm max} \, | L_z |$ is the maximal 
value of the $z$-component of the quark orbital angular momentum 
in the LF wavefunction~\cite{Soft_wall2aa,Soft_wall2b}. 
In particular, we have the values
$L=0$ for $J^P = 0^-$ states and $L=1$ for $J^P = 0^+$ states.
Notice that $\Delta$ is identified with the twist $\tau$ of the
two-parton states. Later we will show that $\tau$
for meson states is independent of the total angular momentum $J$,  i.e.
$\Delta_J \equiv \tau = 2 + L$. Notice that both actions have the correct
conformal limit for $z \to 0$, where the dilaton field vanishes and
conformal invariance is restored.

In a next step we modify the above forms of the action to obtain expressions
which are more convenient in the applications. First, one can remove the
dilaton field from the overall exponential by a specific redefinition of the
bulk field $S^\pm$ with:
\eq\label{ref_specific}
S^\pm(x,z) = e^{\mp\varphi(z)/2} S(x,z) \,.
\en
In terms of the field $S(x,z)$ the transformed action,
which now is universal for both versions of the soft-wall model,
reads~\cite{Soft_wall2aa,Soft_wall2ab,Soft_wall2ac,Soft_wall2b}:
\eq\label{S_action}
S_0 = \frac{1}{2} \int d^dx dz \sqrt{g} \,
\biggl[ g^{MN} \partial_M S(x,z) \partial_N S(x,z)
- (\mu_0^2 + V_0(z)) S^2(x,z)  \biggr] \,,
\en
where $V_0(z) = e^{-2A(z)} U_0(z)$ with the effective potential
\eq
U_0(z) = \frac{1}{2} \varphi^{\prime\prime}(z)
+ \frac{1}{4} (\varphi^{\prime}(z))^2
+ \frac{d-1}{2}\varphi^{\prime}(z) A^{\prime}(z) \,.
\en
The last expression is identical with
the light-front effective potential found in Ref.~\cite{Soft_wall2b}
for $d=4$, $J=0$ [see Eq.(10)]:
\eq
U_0(z) = \kappa^4 z^2 - 2 \kappa^2 \,.
\en
With Lorentzian signature the action~(\ref{S_action}) is given by
\eq\label{S_hidden}
S_0 &=&  \frac{1}{2} \int d^dx dz \, e^{B_0(z)}
\biggl[ \partial_\mu S(x,z) \partial^\mu S(x,z)
- \partial_z S(x,z) \partial_z S(x,z)
- \Big(e^{2A(z)}\mu_0^2 + U_0(z)\Big) S^2(x,z) \biggr] \,,\nonumber\\
B_0(z) &=& (d-1) \, A(z) \,.
\en
Now we use a Kaluza-Klein expansion
\eq\label{KK_coord}
S(x,z) =
\sum\limits_n \ S_n(x) \ \Phi_{n}(z)
\en
where $n$ is the radial quantum number, $S_n(x)$
is the tower of the Kaluza-Klein (KK) modes dual to
scalar mesons and $\Phi_{n}$ are their extra-dimensional profiles
(wave-functions). We assume a free propagation
of the bulk field along the $d$ Poincar\'e coordinates with four-momentum
$p$, and a constrained propagation along the $(d+1)$-th coordinate $z$
(due to confinement imposed by the dilaton field). On mass-shell
$p^2 = M_{n0}^2$ the profiles $\Phi_{n}(z)$ obey the~EOM
\eq\label{Eq0}
\Big[ - \frac{d^2}{dz^2} - B'_0(z) \frac{d}{dz}
+ e^{2A(z)} \mu_0^2 + U_0(z) \Big] \Phi_{n}(z) = M^2_{n0} \Phi_{n}(z) \,.
\en
Performing the substitution
\eq
\Phi_{n}(z) = e^{-B_0(z)/2} \phi_{n}(z)
\en
we derive the Schr\"odinger-type EOM for
$\phi_{n}(z)$:
\eq\label{Eq1}
\Big[ - \frac{d^2}{dz^2} + \frac{4L^2 - 1}{4z^2} + U_0(z)
\Big] \phi_{n}(z) = M^2_{n0} \phi_{n}(z)
\en
where
\eq
\phi_{n}(z) = \sqrt{\frac{2 \Gamma(n+1)}{\Gamma(n+L+1)}} \ \kappa^{L+1}
\ z^{L+1/2} \ e^{-\kappa^2 z^2/2} \ L_n^{L}(\kappa^2z^2)
\en
and
\eq\label{mass2_S}
M^2_{n0} = 4 \kappa^2 \Big( n + \frac{L}{2} \Big) \,
\en
is the mass spectrum of the scalar field.
Here we use the generalized Laguerre polynomials
\eq
L_n^m(x) = \frac{x^{-m} e^x}{n!}
\, \frac{d^n}{dx^n} \Big( e^{-x} x^{m+n} \Big) \,.
\en
Notice that the normalizable modes $\Phi_{n}(z)$ and
$\phi_{n}(z)$ obey the following
normalization conditions:
\eq\label{Norm_Cond}
\int\limits_0^\infty dz \, e^{B_0(z)} \Phi_{m}(z) \Phi_{n}(z) =
\int\limits_0^\infty dz \, \phi_{m}(z) \phi_{n}(z) = \delta_{mn} \,.
\en
The mode $\Phi_{n}(z)$ has the correct behavior in both the
ultraviolet (UV) and infrared (IR) limits:
\eq
\Phi_{n}(z) \ \sim \ z^{2+L}
\ \ {\rm at \ small} \ z\,, \quad\quad
\Phi_{n}(z) \to 0 \ \ {\rm at \ large} \ z\,.
\en
Using the KK expansion~(\ref{KK_coord}), the EOM~(\ref{Eq0}) and
the normalization condition~(\ref{Norm_Cond}) for the KK profiles
$\Phi_{n}(z)$, the $d+1$-dimensional action for the bulk field reduces
to a $d$-dimensional action for the scalar fields $S_n(x)$
dual to scalar mesons with masses $M_{n0}$:
\eq
S_0^{(d)} = \frac{1}{2} \sum\limits_n \, \int d^dx
\biggl[ \partial_\mu S_n(x) \, \partial^\mu S_n(x)
- M_{n0}^2 S_n^2(x) \biggr] \,.
\en
This last equation is a manifestation of the gauge-gravity duality.
In particular, it explicitly demonstrates
that effective actions for conventional hadrons in $d$ dimensions
can be generated from actions for bulk
fields propagating in extra $d+1$ dimensions. The effect of the
extra-dimension is encoded in the hadronic mass squared $M_n^2$,
which is the solution of the Schr\"odinger equation~(\ref{Eq1})
for the KK profile in extra dimension $\phi_{n}(z)$.

Another important conclusion is that we explicitly showed
that correctly formulated soft-wall models with positive or
negative dilaton profiles provide the same results,
moving the dilaton field into the potential of the
Schr\"odinger-type equation for the bound-state problem.
One can also start with the action [see Eq.~(\ref{S_action})]
where the dilaton is hidden in the effective potential $V(z)$.

The next question concerns the choice of the $z$-dependence of the vielbein
or the warping of the AdS metric, i.e. the choice of the function $A(z)$.
As stressed before, the conformal-invariant limit is restricted to
$A(z)=\log(R/z)$. What happens if the ``conformal function'' $A(z)$
is changed by an adjustable function -- ``warping function'' $A_W(z)$,
breaking the conformal invariance of the line element $ds^2$ ?
It is easy to show that the Schr\"odinger-type EOM for $\phi_{n}(z)$
is modified by an extra potential term $W_0(z)$, which can be expressed
in terms of $A_W(z)$ and $A(z)$ as
\eq
W_0(z) = \mu^2 \biggl[ e^{2 A_W(z)} - e^{2 A(z)} \biggr]
+ \frac{(d-1)^2}{4} \biggl[A^\prime_W(z) - A^\prime(z)\biggr]^2
+ \frac{d-1}{2} \biggl[
A^{\prime\prime}_W(z) - A^{\prime\prime}(z) \biggr]
\,.
\en
It is clear that $W_0(z) \equiv 0$ for $A_W(z) \equiv A(z)$.
Physical applications and consequences
of the warping of the AdS metric have been studied in detail in
Refs.~\cite{Soft_wall3a,Soft_wall3b,Soft_wall3c}.
On the other hand, in order to guarantee that two types of soft-wall
model (``dilaton'' and ``metric'' ones) describe the same physics one
should compensate such a correction
by adding the corresponding potential in the actions: for the actions
$S^\pm_0$ and for $S_0$ where the dilaton is hidden
[see Eq.~(\ref{S_hidden})] such correction is
\eq
\Delta S_0^W = \frac{1}{2} \int d^d x dz \sqrt{g_W} \, e^{\pm\varphi(z)} \,
e^{- 2 A_W(z)} \, W_0(z) S^\pm(x,z) S^\pm(x,z) =
\frac{1}{2} \int d^d x dz \, e^{B_0(z)} \, W_0(z) S^2(x,z)\,,
\en
where $g_W = e^{2 A_W(z) (d+1)}$ and
$S^\pm(x,z) = e^{\mp\varphi(z)/2 + (A(z)-A_W(z))(d-1)/2} \, S(x,z)$.

\subsection{Vector field}

In the case of a vector bulk field $V_M(x,z)$
the actions with positive and negative dilaton are
\eq
S^+_1 = \frac{1}{2}
\int d^d x dz \sqrt{g} \, e^{\varphi(z)}
\biggl[
- \frac{1}{2} g^{M_1M_2} g^{N_1N_2} V_{M_1N_1}^+(x,z) V_{M_2N_2}^+(x,z)
+ \Big(\mu_1^2 + \Delta V_1(z)\Big) g^{MN} V_M^+(x,z) V_N^+(x,z) \biggr]
\en
and
\eq
S^-_1 = \frac{1}{2}
\int d^dx dz \, \sqrt{g} \, e^{-\varphi(z)}
\, \biggl[
- \frac{1}{2} g^{M_1M_2} g^{N_1N_2} V_{M_1N_1}^-(x,z) V_{M_2N_2}^-(x,z)
+ \mu_1^2 g^{MN} V_M^-(x,z) V_N^-(x,z) \biggr],
\en
where $V_{MN} = \partial_M V_N - \partial_N V_M$
is the stress tensor of the vector field.
As in case of the scalar field, there exists the bulk field
redefinition $V_M^\pm = e^{\mp\varphi(z)} V_M^\mp$,
making the two actions equivalent to each other.
The mass of the vector bulk field is given by
\eq
\mu_1^2 R^2  = \mu_0^2 R^2 + d - 1 = (\Delta - 1) (\Delta + 1 - d) \,.
\en
The difference between the two actions is absorbed in the effective
potential $\Delta V_1(z) = e^{-2A(z)} \Delta U_1(z)$, where
\eq
\Delta U_1(z) =  - \varphi^{\prime\prime}(z)
+ (3-d) \varphi^\prime(z) A^\prime(z) \,.
\en
Notice that $\Delta U_1(z) \equiv 0$ for $d=4$,
$A(z) = \log(R/z)$ and $\varphi(z) = \kappa^2 z^2$.
Again, as in the scalar field case, one can remove the dilaton
field from the overall exponential by a specific redefinition
of the bulk field~(\ref{ref_specific}):
\eq
S_1 = \frac{1}{2} \int d^d x dz \sqrt{g} \,
\biggl[
- \frac{1}{2}  g^{M_1M_2} g^{N_1N_2} V_{M_1N_1}(x,z) V_{M_2N_2}(x,z)
+ \Big(\mu_1^2 + V_1(z)\Big) g^{MN} V_M(x,z) V_N(x,z) \biggr]
\en
where $V_1(z) = e^{-2A(z)} U_1(z)$ and with the effective potential
\eq
U_1(z) = - \frac{1}{2} \varphi^{\prime\prime}(z)
+ \frac{1}{4} (\varphi^{\prime}(z))^2
+ \frac{3-d}{2}\varphi^{\prime}(z) A^{\prime}(z) \,.
\en
This last expression is identical with
the light-front effective potential found in Ref.~\cite{Soft_wall2b}
for $d=4$, $J=1$ [see Eq.(10)]:
\eq
U_1(z) = \kappa^4 z^2 \,.
\en
Using standard algebra and restricting ourselves to the axial gauge
$V_z(x,z)=0$, we write down the action in terms of fields with
Lorentz indices:
\eq\label{Lorentz_actionV1}
S_1 &=& \frac{1}{2} \int d^d x dz \, e^{B_1(z)}
\biggl[ - \frac{1}{2} V_{\mu\nu}(x,z) V^{\mu\nu}(x,z)
+ \partial_z V_\mu(x,z) \partial_z V^\mu(x,z)
+ \Big(e^{2A(z)} \mu_1^2 + U_1(z)\Big) V_\mu(x,z) V^\mu(x,z) \biggr]
\,,\nonumber\\
B_1(z) &=& B_0(z) - 2 A(z) = (d-3) \, A(z) \,.
\en
It is convenient to rescale the vector fields by the
boost (total angular momentum) factor  $e^{A(z)}$ as
$V_\mu(x,z) \to  e^{A(z)} \, V_\mu(x,z)$.
For higher-spin states the boost factor is $e^{J A(z)}$
(see next subsection). Then the action takes the form
\eq\label{Lorentz_actionV2}
\hspace*{-.5cm}S_1 &=& \frac{1}{2} \int d^d x dz \, e^{B_0(z)}
\biggl[ - \frac{1}{2} V_{\mu\nu}(x,z) V^{\mu\nu}(x,z)
+ \partial_z V_\mu(x,z) \partial_z V^\mu(x,z)
+ \Big(e^{2A(z)} \mu_0^2 + U_1(z)\Big) V_\mu(x,z) V^\mu(x,z) \biggr] \,.
\en
We restrict to the transverse components $V^{\mu \perp}$
($\partial_\mu V^{\mu \perp} = 0$) and again use the KK expansion
\eq\label{KK_coord_V}
V^{\mu \perp}(x,z) = \sum\limits_n \ V^{\mu}_n(x) \ \Phi_{n}(z),
\en
where $V^{\mu \perp}_n$ is the tower of the KK modes dual to
vector mesons and $\Phi_{n}$ are their extra-dimensional profiles
(wave-functions). These coincide with the profiles
of scalar mesons, i.e. are independent on total angular momentum $J$.
In the case of vector mesons the EOM for the profile
function $\Phi_{n}$ is given by
\eq\label{Eq0v}
\Big[ - \frac{d^2}{dz^2} - B'_0(z) \frac{d}{dz}
+ e^{2A(z)} \mu_0^2 + U_1(z) \Big] \Phi_{n}(z) = M^2_{n1} \Phi_{n}(z) \,.
\en
Performing the substitution
\eq
\Phi_{n}(z) = e^{-B_0(z)/2} \phi_{n}(z)
\en
we derive the Schr\"odinger-type EOM for
$\phi_{n}(z)$:
\eq\label{Eq1v}
\Big[ - \frac{d^2}{dz^2} + \frac{4L^2 - 1}{4z^2} + U_0(z)
\Big] \phi_{n}(z) = M^2_{n1} \phi_{n}(z)
\en
where
\eq\label{mass2_V}
M^2_{n1} = 4 \kappa^2 \Big( n + \frac{L}{2} + \frac{1}{2} \Big) \,.
\en
One can see that the EOMs for $\Phi_{n}$ and $\phi_n$ in the case of vector
mesons are different from the analogous EOMs for scalar mesons
by the effective dilaton potential $U_1(z) = U_0(z) + 2 \kappa^2$
for $\varphi(z) = \kappa^2 z^2$ and $A(z)=\log(R/z)$,
which generates the shift
in the mass spectrum: $M^2_{n1} = M_{n0}^2 + 2 \kappa^2$.

As in the case of scalar mesons (using the KK expansion, EOM for the wave
functions) we derive the d-dimensional action for the vector mesons
with masses $M_{n1}^2$ from the higher-dimensional
$d+1$ action in terms of their holographic analogues:
\eq
S_1^{(d)} = \sum\limits_n \, \int d^dx
\biggl[ - \frac{1}{4} V_{\mu\nu, n}(x) \, V^{\mu\nu}_n(x)
+ \frac{M_{n1}^2}{2} V_{\mu, n}(x) V^\mu_n(x) \biggr] \,.
\en
Notice that a possible warping of the metric gives the following correction
to the effective potential
\eq
W_1(z) = \mu^2 \biggl[ e^{2 A_W(z)} - e^{2 A(z)} \biggr]
+ \frac{(d-3)^2}{4} \biggl[ A^\prime_W(z) - A^\prime(z) \biggr]^2
+ \frac{d-3}{2} \biggl[ A^{\prime\prime}_W(z)
- A^{\prime\prime}(z) \biggr] \,.
\en
Again in analogy with scalar mesons, such warping can be compensated
by adding the corresponding potential in the action for the
vector bulk field.

Finally, we would like to mention the results of Ref.~\cite{SWplus},
where the sign of the dilaton field was checked in order to fulfil
different constraints going beyond the bound-state problem,
e.g. absence of a massless state in the vector
channel and a related problem --- $q^2$ behavior of the vector-vector
correlator function $\Pi_V(-q^2)$ consistent with QCD, correct
normalization at $q^2=0$ of the vector bulk-to-boundary propagator
corresponding to the electromagnetic current. The conclusion was that
this requirement fix the sign of the dilaton profile:
it should be negative (in our notations). It was shown there that the
bulk-to-boundary propagator of the vector field $V(q,z)$ satisfying
all criteria is given by~\cite{SWplus}:
\eq\label{Vector_bulk}
V(q,z) = \Gamma\Big(1 - \frac{q^2}{4\kappa^2}\Big)
\, U\Big(-\frac{q^2}{4\kappa^2},0,\kappa^2 z^2\Big) \,.
\en
This is related to the Fourier transform of the transverse massless ($\mu_1=0$)
vector bulk field
\eq
V^-_{\mu \perp}(x,z) = \int \frac{d^d q}{(2\pi)^d} e^{-iqx} V_\mu(q) V(q,z)
\en
entering the action with a negative dilaton profile:
\eq
S^-_1 = - \frac{1}{4}
\int d^dx dz \, \sqrt{g} \, e^{-\varphi(z)}
\, g^{M_1M_2} g^{N_1N_2} V_{M_1N_1}^-(x,z) V_{M_2N_2}^-(x,z)
\en
and obeying the following EOM:
\eq
  \partial_z \biggl( \frac{e^{-\varphi(z)}}{z} \partial_z V(q,z) \biggr)
+ q^2 \, \frac{e^{-\varphi(z)}}{z} V(q,z) = 0 \,.
\en
In other words, the correctly defined bulk-to-boundary
propagator of the vector field $V(q,z)$ is identified with the
one obtained in the version with the negative dilaton.
Let us remind that this is different from the bound-state problem
where the KK profile $\Phi_n(z)$ is independent on the dilaton sign.
Therefore, the version with a negative dilaton
can be applied without any restriction for both bound state and scattering
problem and we certainly prefer this realization of soft-wall model.
For illustration we remind the result for the
pion electromagnetic form factor $F_\pi(Q^2)$ in the Euclidean region
calculated with negative dilaton profile~\cite{Soft_wall2ab,Soft_wall5b}:
\eq
F_\pi(Q^2) = \int\limits_0^\infty \frac{dz}{z^3} e^{-\varphi(z)}
V(Q,z) \Phi_0^2(z) = \frac{1}{\displaystyle{1 + \frac{Q^2}{4\kappa^2}}} \,,
\en
where $\Phi_0(z) = \kappa z^2 \sqrt{2}$ is the wave function for pion
with $n = L = 0$ and $V(Q,z)$ is the holographic analogue of electromagnetic
field given by (\ref{Vector_bulk}). It is clear that the change of the sign
of the dilaton profile leads to divergence of the pion form factor.
The similar situations is for the $\rho$-meson form factor~\cite{Soft_wall4a}
and for the nucleon form factors~\cite{Soft_wall7,Soft_wall9}. 

\subsection{Higher $J$ boson fields}

In this section we consider bulk boson fields with higher values of
$J \ge 2$. This problem, in the context of hard- and soft-wall models, has
been considered before in
Refs.~\cite{Hard_wall2,Soft_wall1,Soft_wall2aa,Soft_wall2ab,Soft_wall2ac,%
Soft_wall2b,Soft_wall3b,Soft_wall8}.
In particular, it was shown that the soft-wall model
reproduces the Regge-behavior of the mesonic mass spectrum
$M_{nJ}^2 \sim n + J$. Here, extending our results for scalar and
vector fields, we show that the bound-state problem is independent on
the sign of the dilaton profile.

We describe a bosonic spin-$J$ field $\Phi_{M_1 \cdots M_J}(x,z)$
by a symmetric, traceless tensor, satisfying the conditions
\eq
\nabla^{M_1}  \Phi_{M_1M_2 \cdots M_J} = 0\,, \quad\quad
g^{M_1M_2}  \Phi_{M_1M_2 \cdots M_J} = 0\,.
\en
The actions for the bulk field $\Phi_J$ with positive and negative
dilatons are~\cite{Soft_wall2aa,Soft_wall2ab,Soft_wall2ac,Soft_wall2b}
\eq\label{SW_JPlus}
S^+_J &=& \frac{(-)^J}{2}
\int d^dx dz \, \sqrt{g} \, e^{\varphi(z)} \,
\biggl[ g^{MN} g^{M_1N_1} \cdots g^{M_JN_J}
\nabla_M\Phi_{M_1 \cdots M_J}^+(x,z)
\nabla_N\Phi_{N_1 \cdots N_J}^+(x,z) \nonumber\\
&-& \mu_J^2 \, g^{M_1N_1} \cdots g^{M_JN_J}
\Phi_{M_1 \cdots M_J}^+(x,z)\Phi_{N_1 \cdots N_J}^+(x,z) \biggr]
\en
and~\cite{Soft_wall8}
\eq\label{SW_JMinus}
S^-_J &=& \frac{(-)^J}{2}
\int d^dx dz \, \sqrt{g} \, e^{-\varphi(z)} \,
\biggl[ g^{MN} g^{M_1N_1} \cdots g^{M_JN_J}
\nabla_M\Phi_{M_1 \cdots M_J}^-(x,z)
\nabla_N\Phi_{N_1 \cdots N_J}^-(x,z) \nonumber\\
&-& \Big(\mu_J^2 + \Delta V_J(z)\Big) \, g^{M_1N_1} \cdots g^{M_JN_J}
\Phi_{M_1 \cdots M_J}^-(x,z)\Phi_{N_1 \cdots N_J}^-(x,z) \biggr]
\,.
\en
Here $\nabla_M$ is the covariant derivative
with respect to AdS coordinates, which is defined as
\eq
\nabla_M \Phi_{M_1 \cdots M_J} =
\partial_M \Phi_{M_1 \cdots M_J}
- \Gamma^N_{M_1M} \Phi_{N M_1 \cdots M_J}
- \cdots - \Gamma^N_{M_JM} \Phi_{M_1 \cdots M_{J-1}N}
\en
where
\eq
\Gamma^{K}_{MN} = \frac{1}{2} g^{KL}
\Big(
  \frac{\partial g_{LM}}{\partial x^N}
+ \frac{\partial g_{LN}}{\partial x^M}
- \frac{\partial g_{MN}}{\partial x^L}
\Big)
\en
is the affine connection.

In Refs.~\cite{Soft_wall2aa,Soft_wall2ab,Soft_wall2ac,Soft_wall2b,Soft_wall8}
higher spin fields have been considered in a ``weak gravity'' approximation,
restricting the analysis to flat metric and therefore
identifying the covariant derivative with the normal derivative
(i.e. neglecting the affine connection).
First, we review these results and then consider the
general case with covariant derivatives. In the following we
call the scenario with normal derivatives scenario I
and the scenario with covariant derivatives scenario II.

In scenario I the bulk mass is given by
by~\cite{Soft_wall2aa,Soft_wall2ab,Soft_wall2ac,Soft_wall2b,Soft_wall8}
\eq
\mu_J^2 R^2 = (\Delta - J) (\Delta + J - d) \,
\en
which is fixed by the behavior of bulk fields $\Phi_J$ near the ultraviolet
boundary $z=0$. The potential  $\Delta V_J(z) = e^{-2A(z)} \Delta U_J(z)$
is given by
\eq
\Delta U_J(z) =  \varphi^{\prime\prime}(z)
+ (d-1-2J) \, \varphi^\prime(z) \, A^\prime(z) \,.
\en
Notice that both quantities $\mu_J^2$ and $\Delta U_J(z)$
are generalizations of the scalar ($J=0$) and vector ($J=1$) cases
considered before. In particular, they are related to
those for the scalar field as follows:
\eq
\mu_J^2 R^2  = \mu_0^2 R^2 + J (d - J)\,, \quad\quad
\Delta U_J(z) = \Delta U_0(z) - 2J \varphi^\prime(z) A^\prime(z) \,.
\en
As before the two actions can be reduced to the action with
a dilaton hidden in an additional potential term using the transformation
\eq\label{ref_specific_J}
  \Phi_{M_1 \cdots M_J}^{\pm}(x,z)
= e^{\mp\varphi(z)/2} \Phi_{M_1 \cdots M_J}(x,z) \,.
\en
Then the action takes the form
\eq
S_J &=& \frac{(-)^J}{2}
\int d^d x dz \sqrt{g}
\biggl[ g^{MN} g^{M_1N_1} \cdots g^{M_JN_J} \,
\partial_M\Phi_{M_1 \cdots M_J}(x,z) \,
\partial_N\Phi_{N_1 \cdots N_J}(x,z) \nonumber\\
&-& (\mu_J^2 + V_J(z)) \, g^{M_1N_1} \cdots g^{M_JN_J}
\Phi_{M_1 \cdots M_J}(x,z) \, \Phi_{N_1 \cdots N_J}(x,z) \biggr]
\en
where $V_J(z) = e^{-2A(z)} U_J(z)$, and with the effective potential
\eq
U_J(z) = \frac{1}{2} \varphi^{\prime\prime}(z)
+ \frac{1}{4} (\varphi^{\prime}(z))^2
+ \frac{d-1-2J}{2}\varphi^{\prime}(z) A^{\prime}(z) \,.
\en
This last expression is identical with
the light-front effective potential found in Ref.~\cite{Soft_wall2b}
for $d=4$ and arbitrary $J$ [see Eq.(10)]:
\eq
U_J(z) = \kappa^4 z^2 + 2 \kappa^2 (J - 1) \,.
\en
Using standard algebra and restricting to the axial gauge
$\Phi_{\cdots z \cdots}(x,z)=0$, we write down the action
in terms of fields with Lorentz indices:
\eq\label{Lorentz_actionPhiJ1}
S_J &=& \frac{(-)^J}{2} \int d^d x dz \, e^{B_J(z)}
\biggl[
  \partial_\mu\Phi_{\mu_1 \cdots \mu_J}(x,z)
  \partial^\mu\Phi^{\mu_1 \cdots \mu_J}(x,z)
- \partial_z\Phi_{\mu_1 \cdots \mu_J}(x,z)
  \partial_z\Phi^{\mu_1 \cdots \mu_J}(x,z) \nonumber\\
&-& \Big(e^{2A(z)} \mu_J^2 + U_J(z)\Big) \Phi_{\mu_1 \cdots \mu_J}(x,z)
  \Phi^{\mu_1 \cdots \mu_J}(x,z) \biggr]
\,,\nonumber\\
B_J(z) &=& B_0(z) - 2 J A(z) = (d - 1 - 2J) \, A(z) \,.
\en
It is convenient to rescale the fields by the
boost (total angular momentum) factor  $e^{J A(z)}$ as
$\Phi_{\mu_1 \cdots \mu_J}(x,z) \to  e^{J A(z)} \,
 \Phi_{\mu_1 \cdots \mu_J}(x,z)$.
Then the action takes the form
\eq\label{Lorentz_actionPhiJ2}
\hspace*{-.5cm}S_J &=& \frac{(-)^J}{2} \int d^d x dz \, e^{B_0(z)}
\biggl[
  \partial_\mu\Phi_{\mu_1 \cdots \mu_J}(x,z)
  \partial^\mu\Phi^{\mu_1 \cdots \mu_J}(x,z)
- \partial_z\Phi_{\mu_1 \cdots \mu_J}(x,z)
  \partial_z\Phi^{\mu_1 \cdots \mu_J}(x,z) \nonumber\\
&-& \Big(e^{2A(z)} \mu_0^2 + U_J(z)\Big) \Phi_{\mu_1 \cdots \mu_J}(x,z)
    \Phi^{\mu_1 \cdots \mu_J}(x,z) \biggr] \,.
\en
Doing the KK expansion
\eq\label{KK_coord_PhiJ}
\Phi^{\mu_1 \cdots \mu_J}(x,z)
= \sum\limits_n \ \Phi^{\mu_1 \cdots \mu_J}_n(x) \ \Phi_{n}(z)
\en
we derive an EOM for the profile function $\Phi_{n}$
\eq\label{Eq0J}
\Big[ - \frac{d^2}{dz^2} - B'_0(z) \frac{d}{dz}
+ e^{2A(z)} \mu_0^2 + U_J(z) \Big] \Phi_{n}(z) = M^2_{nJ} \Phi_{n}(z) \,.
\en
We stress again that the profile function $\Phi_n(z)$ is an
universal function independent on $J$. All the $J$ dependence is hidden
in the effective potential $U_J(z)$, which generates the corresponding
$J$-dependence in the mass spectrum $M^2_{nJ}$.

Performing the substitution
\eq
\Phi_{n}(z) = e^{-B_0(z)/2} \phi_{n}(z)
\en
we derive the Schr\"odinger-type EOM for
$\phi_{n}(z)$:
\eq\label{Eq1J}
\Big[ - \frac{d^2}{dz^2} + \frac{4L^2 - 1}{4z^2} + U_J(z)
\Big] \phi_{n}(z) = M^2_{nJ} \phi_{n}(z),
\en
where
\eq\label{mass2_J}
M^2_{nJ} = 4 \kappa^2 \Big( n + \frac{L}{2} + \frac{J}{2} \Big) \,
\en
is the mass spectrum of higher $J$ fields, which generalizes our results
for scalar and vector fields,~(\ref{mass2_S}) and~(\ref{mass2_V}).
At large values of $J$ or $L$ we reproduce the Regge behavior
of the meson mass spectrum:
\eq
M^2_{nJ} \sim  n + J  \,.
\en
Finally, using the KK expansion and the EOMs for the wave functions,
we derive the $d$-dimensional action for mesons with total angular
momentum $J \ge 2$ and masses $M^2_{nJ}$:
\eq
S_J^{(d)} = \frac{(-)^J}{2} \sum\limits_n \, \int d^dx
\biggl[ \partial_\mu\Phi_{\mu_1 \dots \mu_J, n}(x) \,
\partial^\mu\Phi^{\mu_1 \dots \mu_J}_n(x)
- M_{nJ}^2 \Phi_{\mu_1 \cdots \mu_J, n}(x)
\Phi^{\mu_1 \cdots \mu_J}_n(x) \biggr] \,.
\en
As in the cases of scalar and vector fields, the warping of the metric can
give the following correction to the effective potential
\eq
W_J(z) = \mu^2 \biggl[ e^{2 A_W(z)} - e^{2 A(z)} \biggr]
+ \frac{(d-1-2J)^2}{4} \biggl[ A^\prime_W(z) - A^\prime(z) \biggr]^2
+ \frac{d-1-2J}{2} \biggl[ A^{\prime\prime}_W(z)
- A^{\prime\prime}(z) \biggr] \,,
\en
which can be compensated by adding the following term in the effective
actions $S_J^\pm$, $S_J$:
\eq
\Delta S_J^W &=&
\frac{(-)^J}{2} \int d^d x dz \sqrt{g_W} \, e^{\pm\varphi(z)} \,
e^{-2A_W(z)} \, W_J(z) \, \Phi_{M_1 \cdots M_J}^\pm(x,z) \,
\Phi^{M_1 \cdots M_J, \pm}(x,z) \nonumber\\
&=&\frac{(-)^J}{2} \int d^d x dz \, e^{A_W(z)(d-1-2J)} \, e^{\pm\varphi(z)} \,
W_J(z) \, \Phi_{\mu_1 \cdots \mu_J}^\pm(x,z) \,
\Phi^{\mu_1 \cdots \mu_J, \pm}(x,z) \nonumber\\
&=&
\frac{(-)^J}{2} \int d^d x dz \, e^{B_J(z)} \, W_J(z) \,
\Phi_{\mu_1 \cdots \mu_J}(x,z) \, \Phi^{\mu_1 \cdots \mu_J}(x,z) \,.
\en
where $\Phi_{\mu_1 \cdots \mu_J}^\pm(x,z)
= e^{\mp\varphi(z)/2 + (A(z)-A_W(z))(d-1-2J)/2} \,
 \Phi_{\mu_1 \cdots \mu_J}(x,z)$.

Now we consider scenario II, i.e. without truncation
of covariant derivatives. The gauge-invariant actions for
the totally symmetric higher spin boson fields have been considered
e.g. in Refs.~\cite{Buchbinder:2001bs}.
In this case the bulk mass is fixed by gauge invariance, and given by
\eq\label{mass_JG}
\mu_J^2 R^2 = J^2 + J (d-5) + 4 - 2d \,.
\en
This mass leads to the following results for the scaling of the KK profiles:
$\Phi_{n}(z) \ \sim \ z^{2+J}$ at $z \to 0$, and their masses are
$M^2_{nJ} = 4 \kappa^2 ( n + J )$, which are acceptable only
for the limiting cases $J=L$ and $J\to \infty$.
Notice that the soft-wall actions~(\ref{SW_JPlus} )and~(\ref{SW_JMinus})
are obtained from gauge-invariant actions for totally symmetric higher
spin boson fields~\cite{Buchbinder:2001bs}
via the introduction of the dilaton field, which breaks conformal and gauge
invariance. Therefore, it is not necessary to use the bulk mass
given by Eq.~(\ref{mass_JG}). In particular, in order to get correct
scaling for the KK profile $\Phi_{n}(z) \ \sim \ z^{2+L}$ and their masses
given by Eq.~(\ref{mass2_J}), we should use
\eq\label{mass_JG2}
\mu_J^2 R^2 = ( \Delta - J ) ( \Delta + J - d) - J
= \mu_0^2 R^2 + J (d - 1 - J) \,.
\en
In this case scenario II is fully equivalent to scenario I.

\section{Fermionic case}

\subsection{Spin $1/2$ fermions}

In the fermion case we first
consider the low-lying $J=1/2$ modes $\Psi_\pm(x,z)$
(here the index $\pm$ corresponds again to scenarios with
positive/negative dilaton profiles, respectively.)
The actions with positive and negative dilaton
read ~\cite{Soft_wall6,Soft_wall7,Soft_wall9}
\eq
S^\pm_{1/2} &=&  \int d^dx dz \, \sqrt{g} \, e^{\pm\varphi(z)} \,
\biggl[ \frac{i}{2} \bar\Psi^\pm(x,z) \epsilon_a^M \Gamma^a
{\cal D}_M \Psi^\pm(x,z)
- \frac{i}{2} ({\cal D}_M\Psi^\pm(x,z))^\dagger \Gamma^0
\epsilon_a^M \Gamma^a \Psi^\pm(x,z) \nonumber\\
&-& \bar\Psi^\pm(x,z) \Big(\mu + V_F(z)\Big) \Psi^\pm(x,z) \biggr] \,.
\en
The quantity $\mu$ is the bulk fermion mass with $m = \mu R = \Delta - d/2$,
where $\Delta$ is the dimension of the baryon interpolating
operator, which is related with the scaling dimension $\tau = 3 + L$
as $\Delta = \tau + 1/2$.  For $J=1/2$ we have two baryon multiplets
$J^P = 1/2^+$ for $L=0$ and $J^P = 1/2^-$ for $L=1$.
$V_F(z) = \varphi(z)/R$ is the dilaton field dependent effective potential.
Its presence is necessary due to the following reason.
When the fermionic fields are rescaled
\eq
\Psi^\pm(x,z) = e^{\mp\varphi(z)/2} \Psi(x,z),
\en
the dilaton field is removed from the overall exponential.
In terms of the field $\Psi(x,z)$, the modified action,
which is universal for both versions of the soft-wall model, reads as
\eq\label{S_F}
S_{1/2} = \int d^dx dz \, \sqrt{g} \,
\biggl[ \frac{i}{2} \bar\Psi(x,z) \epsilon_a^M \Gamma^a {\cal D}_M
\Psi(x,z) - \frac{i}{2} ({\cal D}_M\Psi(x,z))^\dagger \Gamma^0
\epsilon_a^M \Gamma^a \Psi(x,z)
- \bar\Psi(x,z) \Big(\mu + V_F(z)\Big) \Psi(x,z) \biggr] \,.
\en
The form of the potential $V_F(z)$ is constrained in order to
get solutions of the EOMs for fermionic KK modes
of left and right chirality, and the correct asymptotics of the nucleon
electromagnetic form factors at large
$Q^2$~\cite{Soft_wall6,Soft_wall7,Soft_wall9}. Note that a possible
warping of the AdS metric is irrelevant in the fermionic case , because it
can be absorbed in the bulk fermion field upon their rescaling, just like
we removed the exponential prefactor containing the dilaton field.

The covariant derivative for the spin $J=1/2$ field is obtained from
the normal derivative by adding the spin connection term:
$\omega_M^{ab} = A^\prime(z) \,
(\delta^a_z \delta^b_M - \delta^b_z \delta^a_M)$:
\eq
{\cal D}_M = \partial_M - \frac{1}{8} \omega_M^{ab} \, [\Gamma_a, \Gamma_b]
\en
where $\Gamma^a=(\gamma^\mu, - i\gamma^5)$ are the Dirac matrices.

The action in terms of the field with Lorentz indices is:
\eq\label{S_F2}
S_{1/2} = \int d^dx dz \, e^{A(z) d} \,
\bar\Psi(x,z)
\biggl[ i\not\!\partial + \gamma^5\partial_z
+ \frac{d}{2} A^\prime(z) \gamma^5
-  \frac{e^{A(z)}}{R} \Big(m + \varphi(z)\Big)  \biggr] \Psi(x,z),
\en
where the Dirac field satisfies the following
EOM~\cite{Soft_wall6,Soft_wall7,Soft_wall9}:
\eq
\biggl[ i\not\!\partial + \gamma^5\partial_z
+ \frac{d}{2} A^\prime(z) \gamma^5
-  \frac{e^{A(z)}}{R} \Big(m + \varphi(z)\Big)  \biggr] \Psi(x,z) = 0 \,,
\en
with $\not\!\partial = \gamma^\mu \, \partial_\mu$.
For the conformal-invariant
metric with $A(z) = \log(R/z)$ we get
\eq
\biggl[ \ \!\!{iz}\not\!\partial + \gamma^5z\partial_z
- \frac{d}{2} \gamma^5
-  m - \varphi(z) \biggr] \Psi(x,z) = 0 \,.
\en
Based on these solutions the fermionic action should be extended by
an extra term in the ultraviolet boundary (see details in
Refs.~\cite{Soft_wall7,Henningson:1998cd}).
Here we review the derivation of the EOMs for the KK modes
dual to the left- and right-chirality spinors in the soft-wall
model~\cite{Soft_wall6,Soft_wall7,Soft_wall9}.
First we expand the fermion field in left- and right-chirality
components:
\eq
\Psi(x,z) = \Psi_L(x,z) + \Psi_R(x,z)\,, \quad
\Psi_{L/R} = \frac{1 \mp \gamma^5}{2} \Psi \,, \quad
\gamma^5 \Psi_{L/R} = \mp \Psi_{L/R} \,.
\en
Then we perform a KK expansion for the $\Psi_{L/R}(x,z)$ fields:
\eq
\Psi_{L/R}(x,z) = \sum\limits_n
\ \Psi_{L/R}^n(x) \ F^n_{L/R}(z) \,.
\en
The KK modes $F^n_{L/R}(z)$ satisfy
the two coupled one-dimensional EOMs~\cite{Soft_wall6,Soft_wall7,Soft_wall9}:
\eq
\biggl[\partial_z \pm \frac{e^{A}}{R} \, \Big(m+\varphi\Big)
+ \frac{d}{2} A^\prime \biggr] F^n_{L/R}(z) = \pm M_n
F^n_{R/L}(z) \,,
\en
where $M_n$ is the mass of baryons with $J=1/2$.
After straightforward algebra one can obtain decoupled
EOMs:
\eq
\biggl[ -\partial_z^2 - d A^\prime \partial_z
+ \frac{e^{2A}}{R^2} (m+\varphi)^2
\mp \frac{e^{A}}{R} \Big(A^\prime (m+\varphi) + \varphi^\prime\Big)
- \frac{d^2}{4} A^{\prime 2} - \frac{d}{2} A^{\prime\prime}
\biggr] F^n_{L/R}(z) = M_n^2  F^n_{L/R}(z) \,.
\en
After the substitution
\eq
F^n_{L/R}(z) = e^{-  A(z) \cdot d/2} \, f^n_{L/R}(z)
\en
we derive the Schr\"odinger-type EOM for $f^n_{L/R}(z)$
\eq
\biggl[ -\partial_z^2
+ \frac{e^{2A}}{R^2} (m+\varphi)^2
\mp \frac{e^{A}}{R} \Big(A^\prime (m+\varphi) +
\varphi^\prime\Big) \biggr] f^n_{L/R}(z) = M_n^2 \, f^n_{L/R}(z) \,.
\en
For $A(z)=\log(R/z)$ and $\varphi(z)=\kappa^2 z^2$ we get
\eq
\biggl[ -\partial_z^2
+ \kappa^4 z^2 + 2 \kappa^2 \Big(m \mp \frac{1}{2} \Big)
+ \frac{m (m \pm 1)}{z^2} \biggr] f^n_{L/R}(z) = M_n^2 \, f^n_{L/R}(z),
\en
where
\eq
f^n_{L}(z) &=& \sqrt{\frac{2\Gamma(n+1)}{\Gamma(n+m+3/2)}} \ \kappa^{m+3/2}
\ z^{m+1} \ e^{-\kappa^2 z^2/2} \ L_n^{m+1/2}(\kappa^2z^2) \,, \\
f^n_{R}(z) &=& \sqrt{\frac{2\Gamma(n+1)}{\Gamma(n+m+1/2)}} \ \kappa^{m+1/2}
\ z^{m} \ e^{-\kappa^2 z^2/2} \ L_n^{m-1/2}(\kappa^2z^2)
\en
with $\int\limits_0^\infty dz \, f^m_{L/R}(z) f^n_{L/R}(z) = \delta_{mn}$
and
\eq
M_n^2 = 4 \kappa^2 \Big( n + m + \frac{1}{2} \Big) \,.
\en
For $d=4$ we have $m = L + 3/2$ and, therefore,
\eq
f^n_{L}(z) &=& \sqrt{\frac{2\Gamma(n+1)}{\Gamma(n+L+3)}} \ \kappa^{L+3}
\ z^{L+5/2} \ e^{-\kappa^2 z^2/2} \ L_n^{L+2}(\kappa^2z^2) \,, \\
f^n_{R}(z) &=& \sqrt{\frac{2\Gamma(n+1)}{\Gamma(n+L+2)}} \ \kappa^{L+2}
\ z^{L+3/2} \ e^{-\kappa^2 z^2/2} \ L_n^{L+1}(\kappa^2z^2)
\en
and
\eq
M_n^2 = 4 \kappa^2 \Big( n + L + 2 \Big) \,,
\en
where $L=0, 1$ for $J=1/2$ fermions.
One can see that
the functions $F^n_{L/R}(z)$ have the correct scaling behavior
for small $z$
\eq
F^n_{L}(z) \sim z^{9/2+L}\,, \quad\quad
F^n_{R}(z) \sim z^{7/2+L}\,
\en
and vanish at large $z$ (confinement).
As in the bosonic case, integration over the holographic coordinate $z$
gives a $d$-dimensional action for the fermion field
$\Psi^n(x) = \Psi^n_L(x) + \Psi^n_R(x)$:
\eq
S_{1/2}^{(d)} = \sum\limits_n \, \int d^dx
\bar\Psi^n(x) \biggl[ i \not\!\partial - M_{n} \biggr] \Psi^n(x) \,.
\en

\subsection{$J=3/2$ and higher $J \ge 5/2$ fermion fields}

Extension of our formalism to $J=3/2$ and higher spin states $J \ge 5/2$
is straightforward. In particular, for $J=3/2$ states we should
construct the action in terms of spinor-vector fields $\Psi_M$,
where $M$ is the AdS index:
\eq
S^\pm_{3/2} &=&  \int d^dx dz \, \sqrt{g} \, e^{\pm\varphi(z)} \,
g^{KN} \,
\biggl[ \frac{i}{2} \bar\Psi^\pm_{K}(x,z) \epsilon_a^M \Gamma^a
{\cal D}_M \Psi^\pm_N(x,z)
- \frac{i}{2} ({\cal D}_M\Psi^\pm_K(x,z))^\dagger \Gamma^0
\epsilon_a^M \Gamma^a \Psi^\pm_N(x,z) \nonumber\\
&-& \bar\Psi^\pm_K(x,z) \Big(\mu + V_F(z)\Big)
\Psi^\pm_N(x,z) \biggr]
\en
and ${\cal D}_M$ is the covariant derivative acting on
the spinor-vector field $\Psi_N$ as
\eq
{\cal D}_M \Psi_N = \partial_M \Psi_N - \Gamma^K_{MN} \Psi_K
-  \frac{1}{8} \omega_M^{ab} [\Gamma_a, \Gamma_b] \Psi_N \,.
\en
Notice that the spin $\omega_M^{ab}$ and affine $\Gamma^K_{MN}$
connections are related as
\eq
\omega_M^{ab} = \epsilon_K^a \Big( \partial_M\epsilon^{Kb}
+ \epsilon^{Nb} \, \Gamma_{MN}^K \Big) \,.
\en
After removing the dilaton field from the exponential prefactor,
doing the boost of the spin-vector field
$\Psi_M(x,z) \to e^{A(z)} \Psi_M(x,z)$, and restricting to the
axial gauge $\Psi_z(x,z) = 0$, we derive the action in terms of fields
with Lorentz indices:
\eq\label{S_F32}
S_{3/2} = \int d^dx dz \, e^{A(z) d} \,
\bar\Psi^\mu(x,z)
\biggl[ i\not\!\partial + \gamma^5\partial_z
+ \frac{d}{2} A^\prime(z) \gamma^5
-  \frac{e^{A(z)}}{R} \Big(m  +  \varphi(z)\Big)  \biggr] \Psi_\mu(x,z) \,.
\en
Then we proceed in analogy with the $J=1/2$ case, and derive the
same $L$ dependent and $J$ independent EOM, which is consistent
with the results of Ref.~\cite{Soft_wall3b}.

Finally, the actions for higher spin $J \ge 5/2$ fermions
with positive and negative dilaton are written as
\eq
\hspace*{-.5cm}
S^\pm_{J} &=&  \int d^dx dz \, \sqrt{g} \, e^{\pm\varphi(z)} \,
g^{K_1N_1} \cdots g^{K_{J-1/2}N_{J-1/2}} \,
\biggl[ \frac{i}{2} \bar\Psi^\pm_{K_1 \cdots K_{J-1/2}}(x,z)
\epsilon_a^M \Gamma^a {\cal D}_M \Psi^\pm_{N_1 \cdots N_{J-1/2}}(x,z)
\nonumber\\
&-& \frac{i}{2}
({\cal D}_M\Psi^\pm_{K_1 \cdots K_{J-1/2}}(x,z))^\dagger
\Gamma^0 \epsilon_a^M \Gamma^a \Psi^\pm_{N_1 \cdots N_{J-1/2}}(x,z)
- \bar\Psi^\pm_{K_1 \cdots K_{J-1/2}}(x,z) \Big(\mu + V_F(z)\Big)
\Psi^\pm_{N_1 \cdots N_{J-1/2}}(x,z) \biggr] \,,
\en
where the covariant derivative ${\cal D}_M$ acting on spin-tensor
field $\Psi^\pm_{N_1 \cdots N_{J-1/2}}$ is
\eq
{\cal D}_M \Psi_{N_1 \cdots N_{J-1/2}} &=&
\partial_M \Psi_{N_1 \cdots N_{J-1/2}}
- \Gamma^K_{MN_1} \Psi_{K N_2 \cdots N_{J-1/2}}
- \cdots - \nonumber\\
&-& \Gamma^K_{MN_{J-1/2}} \Psi_{N_1 \cdots N_{J-3/2} K}
- \frac{1}{8} \omega_M^{ab} [\Gamma_a, \Gamma_b]
\Psi_{N_1 \cdots N_{J-1/2}} \,.
\en
As before, for the $J=1/2, 3/2$ cases we remove the dilaton field from
the exponential prefactor, perform the boost of the spin-tensor field
and restrict ourselves to the axial gauge. Then we derive the action in
terms of fields with Lorentz indices:
\eq\label{S_FJ}
S_{J} = \int d^dx dz \, e^{A(z) d} \,
\bar\Psi^{\mu_1 \cdots \mu_{J-1/2}}(x,z)
\biggl[ i\not\!\partial + \gamma^5\partial_z
+ \frac{d}{2} A^\prime(z) \gamma^5
-  \frac{e^{A(z)}}{R} \Big(m  +  \varphi(z)\Big)  \biggr]
\Psi_{\mu_1 \cdots \mu_{J-1/2}}(x,z) \,.
\en
Next, after a straightforward algebra (including the KK expansion),
we derive the same equation of motion for the KK profile and mass
formula as for fermions with lower spins. The action for
physical baryons with higher spins (the result for $J=3/2$ is straightforward)
is written as
\eq
S_{J}^{(d)} = \sum\limits_n \, \int d^dx
\bar\Psi^{\mu_1 \cdots \mu_{J-1/2}, n}(x)
\biggl[ i \not\!\partial - M_{n} \biggr]
\Psi_{\mu_1 \cdots \mu_{J-1/2}}^n(x) \,.
\en
Therefore, the main difference
between the bosonic and fermionic actions is that in the case of bosons the
mass formula depends on the combination $(J+L)/2$, while in the baryon case
it depends only on $L$. Also, in the fermion case the dilaton prefactor
and possible warping of conformal-invariant AdS metric can be easily
absorbed in the field, without the generation of extra potential terms.

\section{Applications}

\subsection{Basic properties of pion and nucleon}

We consider several applications of our soft-wall model with negative
dilaton profile. First, we display the predictions for basic properties
of pions and nucleons with the same value of dilaton
parameter $\kappa = 350$ MeV.

The masses of pion and nucleon are:
\eq
M_\pi = 0\, \ \
M_N = 2 \sqrt{2} \kappa = 990 \ {\rm MeV} \,.
\en
The pion decay constant is given by~\cite{Soft_wall2b}:
\eq
F_\pi = \frac{\sqrt{3}}{8} \kappa = 76 \ {\rm MeV} \,.
\en
The electromagnetic radii of pion and nucleon
are given by the expression (see details in
Refs.~\cite{Soft_wall2ab,Soft_wall5b,Soft_wall7,Soft_wall9}):
\eq
\la r^2 \ra^\pi &=& \frac{12}{M_N^2} \,, \ \ \
\la r^2_E \ra^p  \ = \  \frac{147}{8 M_N^2}
\biggl( 1 + \frac{13}{147} \mu_p \biggr) \,, \ \ \
\la r^2_E \ra^n \ = \  \frac{13}{8 M_N^2} \mu_n\,, \nonumber\\
\la r^2_M \ra^p &=& \frac{177}{8 M_N^2}
\biggl( 1 - \frac{17}{177 \mu_p} \biggr) \,, \ \ \
\la r^2_M \ra^n \ = \  \frac{177}{8 M_N^2} \,,
\en
where $\mu_p$ and $\mu_n$ are the magnetic moments of
nucleons. Here, for convenience, we expressed the dilaton parameter
through the nucleon mass. Using data for $\mu_p = 2.793$ and
$\mu_n = - 1.913$, the results for the slopes compared rather well with
data:
\eq
\la r^2 \ra^\pi &=& 0.476 \ {\rm fm}^2 \ {\rm (our)}\,,
\quad
0.452 \ {\rm fm}^2 \ {\rm (data)}\,, \nonumber\\
\la r^2_E \ra^p &=& 0.910 \ {\rm fm}^2 \ {\rm (our)}\,,
\quad
0.766 \ {\rm fm}^2 \ {\rm (data)}\,, \nonumber\\
\la r^2_E \ra^n &=& - 0.123 \ {\rm fm}^2 \ {\rm (our)}\,,
\quad
- 0.116 \ {\rm fm}^2 \ {\rm (data)}\,, \nonumber\\
\la r^2_M \ra^p &=& 0.849 \ {\rm fm}^2 \ {\rm (our)}\,,
\quad
0.731 \ {\rm fm}^2 \ {\rm (data)}\,, \\
\la r^2_M \ra^n &=& 0.879 \ {\rm fm}^2 \ {\rm (our)}\,,
\quad
0.762 \ {\rm fm}^2 \ {\rm (data)}\,. \nonumber
\en
Other important applications of our approach can be found
in~\cite{Soft_wall8,Soft_wall9}. In particular,
in~\cite{Soft_wall8} we presented a detailed analysis of meson
mass spectrum and decay constants. Moreover,
in Ref.~\cite{Soft_wall9} we did the first calculation of
nucleon generalized parton distributions in AdS/QCD.

\subsection{Baryon mass spectrum}

Here we present the application of our approach to the baryon
spectrum. We remind that the baryon mass spectrum calculated
in our formalism is specified
by the radial quantum number $n$ and orbital angular momentum $L$.
\eq\label{mass_lead}
M_{nL}^2 = 4 \kappa^2 \Big( n + L + 2 \Big) \,.
\en
It means that the states with different spin $S=1/2$ and $S=3/2$
and fixed $L$ are degenerate. As is well-known, this degeneracy is
removed by taking into account a hyperfine (HF) spin-spin
interaction between quarks in the baryon, due to one-gluon
exchange~\cite{Close:bt,Donoghue:dd,Inoue:2004jb,Karliner:2008sv}:
\eq
H_{\mbox{\footnotesize hyp}}
= \frac12 \sum\limits_{i \, < \, j} {\cal H}_{ij} \,
\vec s_i \cdot \vec s_j \, \delta^3(\vec r_i -\vec r_j)
\label{eqn:hyperint}
\en
where $\vec s_i$ is the spin operator acting on the $i$-th quark.
Here, ${\cal H}_{ij}$ is the two-body quark coupling, which includes
a common color factor $-2/3$ and explicitly depends on the flavor of
the constituent quarks through their masses $m_i$ and $m_j$:
\eq
{\cal H}_{ij} \, \sim \, \frac{2}{3} \, \frac{1}{m_i \, m_j} \,.
\en
The use of $SU(6)$ spin-flavor wave functions for the ground-state
baryons $B$ leads to simple relations between the matrix elements
$\la B|H_{\mbox{\footnotesize hyp}}|B \ra$,
which are the perturbative mass shifts due to the HF interaction.

Denoting the contribution from a non-strange quark pair as
${\cal H}_{qq}$ (and similarly for strange quarks with $q$ replaced
by $s$), in the isospin limit the masses of the ground-state baryons
are composed as~\cite{Donoghue:dd}
\eq\label{mass_formula_hyp}
\begin{array}{l}
  M_N         = 3 E_0            - \frac{3}{8} {\cal H}_{qq}\,, \\
  M_{\Lambda} = 2 E_0 +   E_0^s  - \frac{3}{8} {\cal H}_{qq}\,, \\
  M_{\Sigma}  = 2 E_0 +   E_0^s  + \frac{1}{8} {\cal H}_{qq} -
  \frac{1}{2} {\cal H}_{qs}\,, \\
  M_{\Xi}     =   E_0 + 2 E_0^s  - \frac{1}{2} {\cal H}_{qs}
  + \frac{1}{8} {\cal H}_{ss}\,,
 \end{array}
 \hspace*{1cm}
 \begin{array}{l}
  M_{\Delta~}    = 3 E_0           + \frac{3}{8} {\cal H}_{qq} \\
  M_{\Sigma^\ast}= 2 E_0 +   E_0^s + \frac{1}{8} {\cal H}_{qq}
  + \frac14 {\cal H}_{qs} \\
  M_{\Xi^\ast}   =   E_0 + 2 E_0^s + \frac{1}{4} {\cal H}_{qs}
  + \frac18 {\cal H}_{ss} \\
  M_{\Omega~}    =         3 E_0^s + \frac{3}{8} {\cal H}_{ss}
\end{array}
\en
Here $E_0$ and $E_0^s$
are the single particle ground-state energies of the non-strange and
strange quark, respectively.
These mass formulas satisfy the Gell-Mann-Okubo mass relations
\begin{eqnarray}\label{GMO}
&&M_{\Sigma} - M_N = \frac{1}{2} (M_{\Xi}-m_N) +
 \frac{3}{4} (M_{\Sigma}-M_{\Lambda})\,,  \hspace*{.25cm}
M_{\Sigma^\ast} - M_{\Delta} = M_{\Xi^\ast}-M_{\Sigma^\ast} =
M_{\Omega} - M_{\Xi^\ast}
\end{eqnarray}
providing a condition on the matrix elements of the residual
interaction with
${\cal H}_{qq} - {\cal H}_{qs} \simeq {\cal H}_{qs} - {\cal H}_{ss}$.
With the choice $E_0^s - E_0 \simeq 180$ MeV, ${\cal H}_{qq} \simeq 400$ MeV
and ${\cal H}_{qq} - {\cal H}_{qs} \simeq {\cal H}_{qs} - {\cal H}_{ss}
\simeq 150$ MeV,
all the observed mass differences can be approximately reproduced.
Another important piece which contributes to the baryon masses
is the meson cloud (MC) induced mass shift, due to the interaction between
quarks and pseudoscalar $(\pi, K, \eta)$ meson fields. Such
contribution was evaluated using different chiral quark models
(see e.g. detailed discussion in Ref.~\cite{Inoue:2004jb}).
It was proved that MC contribution is negative and is similar
on magnitude for the octet and decuplet states.

Here we suggest a phenomenological formula for the square of the
baryon mass, treating the HF splitting and MC corrections perturbatively
(it means that we restrict to the first-order in such effects).
In particular, we assume the following conjecture for the light baryon masses,
including HF ($\delta M^2_S$) and MC ($\delta M^2_C$) corrections:
\eq\label{mass_full_inv_our}
M^2_B = M^2_{nL} + \delta M^2_S + \delta M^2_C
= 4 \kappa^2 (n + L + \alpha_B S + 2 - \alpha_B - \beta_B) \,.
\en
where $M^2_{nL}$ is the leading term (\ref{mass_lead})
predicted by the soft-wall model, while the terms
$\delta M^2_{\rm S}$ and $\delta M^2$ are:
\eq\label{mass_HF}
\delta M^2_S = 4 \alpha_B \kappa^2 (S-1) \,, \quad
\delta M^2_C = - 4 \beta_B  \kappa^2 \,.
\en
Here $S=1/2$ or $3/2$ is the internal spin,
$\alpha_B$ and $\beta_B$ are free parameters.
Note that at $\alpha_B = 1/2$ and $\beta_B = 3/4$
our empirical formula coincides with the one previously derived by
Brodsky and de T\'eramond~\cite{Soft_wall2ac}, where they
subtracted the constant term from the light-front Hamiltonian matched to
AdS/QCD Hamiltonian:
\eq\label{mass_full_inv_BT}
M^2_B = 4 \kappa^2 (n + L + S/2 + 3/4) \,.
\en
Formula (\ref{mass_full_inv_BT})
was derived in limit of SU(3) flavor invariance. Lets go beyond this
and include SU(3) breaking effects caused by the strange--nonstrange
quark mass difference $\delta_s = m_s - m_q$. Taking into account
the SU(6) algebra for HF couplings [see Eq.~(\ref{mass_formula_hyp})]
and SU(3) shifts $\delta_B$ of the single particle ground-state energies
in the baryon [see Eq.~(\ref{mass_formula_hyp})] we extend the master
formula as
\eq\label{mass_full_inv}
M_B = 2 \kappa \sqrt{n + L + 2 + \alpha_B (S-1) - \beta_B}
+ \delta_B
\en
where
\eq
& &\alpha_N = \alpha_\Lambda = \alpha_D = \alpha\,, \quad
\alpha_\Sigma = \frac{4 r_s - 1}{3} \, \alpha\,,   \quad
\alpha_\Xi = \frac{(4  - r_s) r_s}{3} \, \alpha\,,  \nonumber\\
& &\alpha_{\Sigma^\ast} = \frac{1  + 2 r_s}{3} \, \alpha\,,  \quad
\alpha_{\Xi^\ast} = \frac{(2  + r_s) r_s}{3} \, \alpha\,,  \quad
\alpha_{\Omega} = r_s^2 \alpha
\en
and
\eq
\delta_N = \delta_\Delta = 0\,, \quad
\delta_\Lambda = \delta_\Sigma = \delta_{\Sigma^\ast} = \delta_s\,,
\quad \delta_\Xi = \delta_{\Xi^\ast} = 2 \delta_s\,, \quad
\delta_{\Omega} = 3 \delta_s \,.
\en
Here $r_s = m_q/m_s$.
The SU(3) limit corresponds to the conditions $r_s=1$, $\delta_s = 0$.
For MC correction we use the universal parameter $\beta_B = \beta$.

In our calculation of baryon masses we use 7 free parameters:
the dilaton profile parameter $\kappa$, the couplings $\alpha$ and $\beta$,
the set of constituent quark masses $m_q = m_u = m_d$, $m_s$, $m_c$ and $m_b$.
The parameter $\kappa$ fixes the slopes of the baryon
mass trajectories.
Using data for excited light baryons we fix the value to $\kappa = 500$ MeV,
which is also in accordance with the analysis of Ref.~\cite{Soft_wall2ac}.
Notice that this value is also close to the value of $\kappa = 550$ MeV
found in the study of meson properties in~\cite{Soft_wall8}.
The free parameters $\alpha$ and $\beta$ are fixed by the
nucleon and $\Delta(1232)$ isobar masses, using two constraints
from the baryon mass formula~(\ref{mass_full_inv}):
\eq
M_N^2 = 4 \kappa^2 (2 - \alpha/2 - \beta)\,, \quad
M_\Delta^2 = 4 \kappa^2 (2 + \alpha/2 - \beta)\,,
\en
which lead to
\eq
\alpha = \frac{M_\Delta^2 - M_N^2}{4\kappa^2}\,, \quad
\beta = 2- \frac{M_\Delta^2 + M_N^2}{8\kappa^2} \,.
\en
For $\kappa = 500$ MeV we get $\alpha = 0.636$ and $\beta = 0.800$.
The light quark masses $m_q = 400$ MeV and $m_s = 575$ MeV are fixed from
data for ground-state masses of light hyperons. Finally, the masses of
the charm and bottom quarks $m_c = 1.747$~GeV and $m_b = 5.081$~GeV
are fixed from data on the $\Lambda_c(2286)$ and $\Lambda_b(5620)$ masses. 
Finally we consider the set of parameters
\eq\label{Best_Fit}
& &\kappa = 500 \ {\rm MeV}\,, \hspace*{.5cm}
   \alpha =  0.636\,, \hspace*{.5cm}
   \beta = 0.800\,, \nonumber\\
& &m_q = 400 \ {\rm MeV}\,,  \hspace*{.5cm}
   m_s = 575 \ {\rm MeV}\,,  \hspace*{.5cm}
   m_s = 1.747 \ {\rm GeV}\,,  \hspace*{.5cm}
   m_s = 5.081 \ {\rm GeV} \ \,.
\en
as the best fit.
We want to stress that in our approach we consider constituent quarks.
In particular, in the light sector the masses of the $u,d,s$ quarks
encode spontaneous breaking of chiral symmetry and do not vanish
in the chiral limit (when the corresponding current quark masses vanish).
Moreover, hyperfine splitting effects must be described in terms of constituent
quark masses (the use of current quark masses leads to a divergence of the
hyperfine splitting in the chiral limit). On the other hand, in the context of
chiral quark models (see e.g. discussion in Ref.~\cite{Faessler:2005gd})
it is possible to establish a relation between the constituent and the current
quark masses in the form of a chiral expansion, which is consistent with
low-energy theorems of QCD.
It was shown in~\cite{Faessler:2005gd} that realistic values
for the current masses of $u, d$ and $s$ quarks (found in lattice QCD and fixed
in chiral perturbation theory) correspond to the following values of the
constituent quark masses $m_q = 420$ MeV (for $u,d$-quarks) and
$m_s = 590$ MeV (for the $s$-quark).
These values compare well with the parameters
used in our soft-wall model ($m_q = 400$ MeV and $m_s = 575$ MeV).
In case of heavy quarks we use constituent quark masses which are a
bit larger than the values quoted by the Particle Data Group~\cite{PDG}.
The present set of parameters $(\kappa, m_q, m_s, m_c, m_b)$
is close to those used in the analysis of meson physics~\cite{Soft_wall8}:
\eq
\kappa = 550 \ {\rm MeV}\,, \hspace*{.5cm}
m_q = 420 \ {\rm MeV}\,, \hspace*{.5cm}
m_s = 570 \ {\rm MeV}\,, \hspace*{.5cm}
m_c = 1.6 \ {\rm GeV}\,, \hspace*{.5cm}
m_b = 4.8  \ {\rm GeV}\,.
\en
Also the set of constituent quark masses is very close
to the one used in the Lorentz covariant three-quark model
in a detailed description of exclusive strong, electromagnetic
and weak decays of light and heavy baryons~\cite{LCQM1,LCQM2}:
\eq
m_q = 420 \ {\rm MeV}\,, \hspace*{.5cm}
m_s = 570 \ {\rm MeV}\,, \hspace*{.5cm}
m_c = 1.7 \ {\rm GeV}\,, \hspace*{.5cm}
m_b = 5.2  \ {\rm GeV}\,.
\en
For completeness we indicate the relative error in the calculation
of light and single-heavy baryon masses defined as
$\delta_{\rm err} = |(M_B^{\rm exp} - M_B^{\rm th})/M_B^{\rm exp}|
\cdot 100\%$, where $M_B^{\rm exp}$ and $M_B^{\rm th}$ are the central
values of the respective experimental and theoretical baryon masses.
For the ground states (24 states) we get $\delta_{\rm err} \leq 1 \%$, 
while for the excited ones (60 states) we have $\delta_{\rm err} \leq 5 \%$. 

With the set of parameters given in Eq.~(\ref{Best_Fit}) we indicate our
results for the light and heavy baryon masses in the following
Tables I-VII.
Table I contains the results for the ground states of light baryons.
In Tables II and III we display our results for the excited states of the $N$,
$\Delta$, $\Lambda$, $\Sigma$ and $\Sigma^\ast$  families, with
different values of $n$ and $L$, and compare them with available data. 
In Tables IV and V we present a detailed classification of 
single, double and triple heavy baryons and results for their mass spectrum. 
Also we specify the HF couplings $\alpha_B$ (i.e. the ratio  
$\alpha_B/\alpha$). 
We introduced the notations $r_c = m_q/m_c$ and $r_b = m_q/m_b$. 
Note, we consider the mass spectrum of heavy baryons
containing a single, two and three heavy $b$ or $c$ quarks
using the master formula~(\ref{mass_full_inv}) with the
same parameters of $\alpha$, $\beta$, $m_q$, $m_s$ and $\kappa$.
We compare our results with available
data~\cite{PDG} or with prediction of QCD motivated
relativistic quark models~\cite{Ebert:2004ck,Ebert:2011kk,Martynenko:2007je}.
Single-- and double--heavy baryons are classified by the set
of quantum numbers $(J^P, S_d)$, where $J^P$ is the spin--parity of
the baryon state and $S_d$ is the spin of the light or heavy diquark,
respectively (see details in Ref.~\cite{LCQM1}).
There are two types of light and heavy diquarks -- those with $S_d = 0$
(antisymmetric spin  configuration $[q_1q_2]$) and those with $S_d = 1$
(symmetric spin configuration $\{q_1q_2\}$). Accordingly there are two
$J^{P}=1/2\,^+$  single-- and double--heavy baryon states.
We follow the standard convention and attach
a prime to the $S_d = 1$ states whereas the $S_d = 0$ states are unprimed
in the case of single--heavy baryons, and vice versa
in the case of double--heavy baryons  ---
we attach a prime to the $S_d = 0$ states
whereas the $S_d = 1$ states are unprimed.
Finally, in Tables VI and VII 
we present our results for the mass spectrum of excited states of
single--heavy baryons $\Lambda_Q$, $\Sigma_Q^{(\ast)}$, 
$\Xi_Q^{(\ast)}$ and $\Omega_Q^{(\ast)}$ 
and compare it with the recent prediction of relativistic quark-diquark
model~\cite{Ebert:2011kk}. 

\section{Conclusions}

We performed a systematic analysis of extra-dimensional actions for
bosons and fermions, which give rise to actions for observable hadrons.
Masses of these hadrons are calculated analytically from Schr\"odinger type
equations of motion with a potential which provides confinement of the
Kaluza-Klein (KK) modes in extra $(d+1)$ dimension.
The tower of KK modes with radial quantum number $n$ and total
angular momentum $J$ has direct correspondence to realistic
mesons and baryons living in $d$ dimensions. For such correspondence
the sign of the dilaton profile is irrelevant, because the exponential
prefactor containing the dilaton is finally absorbed in the bulk fields.
On the other hand, the sign of the dilaton profile becomes important
for the definition/calculation of the bulk-to-boundary propagator --
i.e. the Green function describing the evolution of the bulk field from
inside of AdS space to its ultraviolet boundary. The corresponding
sign should be negative in order to fulfill certain constraints. It was 
discussed recently in Refs.~\cite{SWplus}. As application of our approach
we presented detailed analysis of mass spectrum of light and heavy baryons.

\begin{acknowledgments}

The authors thank Stan Brodsky and Guy de T\'eramond for useful discussions
and remarks. This work was supported by Federal Targeted Program "Scientific
and scientific-pedagogical personnel of innovative Russia"
Contract No. 02.740.11.0238, by FONDECYT (Chile) under Grant No. 1100287.
V. E. L. would like to thank Departamento de F\'\i sica y Centro
Cient\'\i fico Tecnol\'ogico de Valpara\'\i so (CCTVal), Universidad
T\'ecnica Federico Santa Mar\'\i a, Valpara\'\i so, Chile for warm
hospitality. A. V. acknowledges the financial support from FONDECYT (Chile)
Grant No. 3100028.

\end{acknowledgments}

\newpage

\begin{table}
\begin{center}
\caption{Mass spectrum of ground-state light baryon in MeV}

\vspace*{.25cm}

\def\arraystretch{1.5}
    \begin{tabular}{|c|c|c|}
      \hline
      Baryon & Our results & Data~\cite{PDG}    \\
\hline
        $N$    &    939  & 939                   \\
  $\Lambda$    &   1114  & 1116                  \\
  $\Sigma $    &   1180  & 1189                  \\
  $\Xi    $    &   1328  & 1322                  \\
  $\Delta $    &   1232  & 1232                  \\
  $\Sigma^\ast$&   1381  & 1385                  \\
  $\Xi^\ast$   &   1533  & 1530                  \\
  $\Omega $    &   1688  & 1672                  \\
      \hline
    \end{tabular}
\end{center}

\begin{center}
\caption{Mass spectrum of $N$ and $\Delta$ families in MeV}

\vspace*{.25cm}

\def\arraystretch{1.5}
\begin{tabular}{|c|c|c|}
      \hline
      Baryon & Our results & Data~\cite{PDG} \\
\hline
$N_{1/2^+}(939)$  &  939  & 939                   \\
$N_{1/2^+}(1440)$ &  1372 & $1440^{+30}_{-20}$   \\
$N_{1/2^+}(1710)$ &  1698 & $1710 \pm 30$        \\
$N_{1/2^+}(1880)$ &  1970 &                    \\
$N_{1/2^+}(2100)$ &  2209 &                    \\
$\Delta_{3/2^+}(1232)$ &  1232 & $1232$   \\
$\Delta_{3/2^+}(1600)$ &  1587 & $1600^{+100}_{-50}$   \\
$\Delta_{3/2^+}(1920)$ &  1876 & $1920^{+50}_{-20}$   \\
      \hline
    \end{tabular}
\end{center}

\begin{center}
\caption{Mass spectrum of $\Lambda$, $\Sigma$ and
$\Sigma^\ast$ families in MeV}

\vspace*{.25cm}

\def\arraystretch{1.5}
    \begin{tabular}{|c|c|c|}
      \hline
Baryon  & Our results  & Data~\cite{PDG}  \\
\hline
$\Lambda_{1/2^+}(1116)$  &  1114 & 1116                  \\
$\Lambda_{1/2^+}(1600)$  &  1547 & $1600^{+100}_{-40}$   \\
$\Lambda_{1/2^+}(1810)$  &  1873 & $1810^{+40}_{-60}$    \\
$\Sigma_{1/2^+}(1189)$   &  1180 & 1189                  \\
$\Sigma_{1/2^+}(1660)$   &  1593 & $1660 \pm 30$         \\
$\Sigma_{1/2^+}(1880)$   &  1910 & 1880                  \\
$\Sigma_{3/2^+}(1385)$   &  1381 & 1385                  \\
$\Sigma_{3/2^+}(1840)$   &  1741 & 1840                  \\
$\Sigma_{3/2^+}(2080)$   &  2033 & 2080                  \\
\hline
    \end{tabular}
\end{center}
\end{table}

\newpage

\begin{table}
\begin{center}
\caption{Classification and mass spectrum (in GeV)
of ground--state single--heavy baryons}

\vspace*{.25cm}
\def\arraystretch{2.5}
\begin{tabular}{|c|c|c|c|c|c|}
\hline
\,\, Baryon \,\,  & \,\, Content \,\, &  \,\, $J^P$ \,\, &
\,\, Ratio $\alpha_B/\alpha$ \,\, & \,\, Our results \,\, & \,\,
Data~\cite{PDG}, Ref.~\cite{Ebert:2011kk}\,\, \\[2mm]
\hline
 $\Lambda_{c}$ & $c[ud]$       & $1/2^+$  & 1
 & 2.286 & $2.286$~\cite{PDG} \\
 $\Xi_{c}$ & $c[sq]$           & $1/2^+$  & $r_s$
 & 2.511 & $2.468$~\cite{PDG} \\
 $\Xi_{c}^{\prime}$ & $c\{sq\}$& $1/2^+$
 & $\displaystyle{\frac{2r_c + 2r_cr_s - r_s}{3}}$
 & 2.613 & $2.576$~\cite{PDG} \\
 $\Sigma_{c}$ & $c\{qq\}$      & $1/2^+$
& $\displaystyle{\frac{4r_c - 1}{3}}$
 & 2.446 & $2.454$~\cite{PDG} \\
 $\Omega_{c}$ & $c\{ss\}$      & $1/2^+$
& $\displaystyle{\frac{(4r_c - r_s)r_s}{3}}$
 & 2.785 & $2.695$~\cite{PDG} \\
 $\Lambda_{b}$ & $b[ud]$       & $1/2^+$  & 1
 & 5.620 & $5.620$~\cite{PDG} \\
 $\Xi_{b}$ & $b[sq]$           & $1/2^+$  & $r_s$
 & 5.845 & $5.791$~\cite{PDG} \\
 $\Xi_{b}^{\prime}$ & $b\{sq\}$& $1/2^+$
& $\displaystyle{\frac{2r_b + 2r_br_s - r_s}{3}}$
 & 5.972 & $5.936$~\cite{Ebert:2011kk}\\
 $\Sigma_{b}$ & $b\{qq\}$      & $1/2^+$
& $\displaystyle{\frac{4r_b - 1}{3}}$
 & 5.809 & $5.809$~\cite{PDG} \\
 $\Omega_{b}$ & $b\{ss\}$      & $1/2^+$
& $\displaystyle{\frac{(4r_b - r_s)r_s}{3}}$
 & 6.139 & $6.071$~\cite{PDG}\\
 $\Xi_{c}^\ast$ & $c\{sq\}$    & $3/2^+$
& $\displaystyle{\frac{r_c + r_s + r_cr_s}{3}}$
 & 2.669 & $2.646$~\cite{PDG}\\
 $\Sigma_{c}^\ast$ & $c\{qq\}$ & $3/2^+$
& $\displaystyle{\frac{2r_c + 1}{3}}$
 & 2.511 & $2.518$~\cite{PDG}\\
 $\Omega_{c}^\ast$ & $c\{ss\}$ & $3/2^+$
& $\displaystyle{\frac{(2r_c + r_s)r_s}{3}}$
 & 2.831 & $2.766$~\cite{PDG}\\
 $\Xi_{b}^\ast$ & $b\{sq\}$    & $3/2^+$
& $\displaystyle{\frac{r_b + r_s + r_br_s}{3}}$
 & 5.991 & $5.963$~\cite{Ebert:2011kk}\\
 $\Sigma_{b}^\ast$ & $b\{qq\}$ & $3/2^+$
& $\displaystyle{\frac{2r_b + 1}{3}}$
 & 5.831 & $5.829$~\cite{PDG}\\
 $\Omega_{b}^\ast$ & $b\{ss\}$ & $3/2^+$
& $\displaystyle{\frac{(2r_b + r_s)r_s}{3}}$
 & 6.155 & $6.088$~\cite{Ebert:2011kk}\\
\hline
\end{tabular}
\end{center}
\end{table}

\newpage
\begin{table}
\begin{center}
\caption{Classification and mass spectrum (in GeV)
of double--heavy and triple--heavy baryons}

\vspace*{.25cm}
\def\arraystretch{2.5}
\begin{tabular}{|c|c|c|c|c|c|}
\hline
\,\, Baryon \,\,  & \,\, Content \,\, &  \,\, $J^P$ \,\, &
\,\, Ratio $\alpha_B/\alpha$ \,\, & \,\, Our results \,\, & \,\,
Data~\cite{PDG}, Refs.~\cite{Ebert:2004ck,Martynenko:2007je}\,\, \\[2mm]
\hline
$\Xi_{cc}$  & $q\{cc\}$ & $1/2^+$ &
$\displaystyle{\frac{(4 - r_c) r_c}{3}}$
& 3.747 & 3.5189~\cite{PDG} \\
$\Xi_{bc}$  & $q\{bc\}$ & $1/2^+$ &
$\displaystyle{\frac{2r_b + 2r_c - r_br_c}{3}}$
& 7.094 & 6.933~\cite{Ebert:2004ck} \\
$\Xi'_{bc}$ & $q[bc]$   & $1/2^+$ & $r_br_c$
& 7.121  & 6.963~\cite{Ebert:2004ck} \\
$\Xi_{bb}$  & $q\{bb\}$ & $1/2^+$
& $\displaystyle{\frac{(4 - r_b) r_b}{3}}$
& 10.442 & 10.202~\cite{Ebert:2004ck} \\
$\Xi_{cc}^\ast$  & $q\{cc\}$ & $3/2^+$
& $\displaystyle{\frac{(2 + r_c) r_c}{3}}$
& 3.814 & 3.727~\cite{Ebert:2004ck} \\
$\Xi_{bc}^\ast$  & $q\{bc\}$ & $3/2^+$
& $\displaystyle{\frac{r_b + r_c + r_br_c}{3}}$
& 7.139 & 6.980~\cite{Ebert:2004ck} \\
$\Xi_{bb}^\ast$  & $q\{bb\}$ & $3/2^+$
& $\displaystyle{\frac{(2 + r_b) r_b}{3}}$
& 10.465 & 10.237~\cite{Ebert:2004ck} \\
$\Omega_{cc}$  & $s\{cc\}$ & $1/2^+$
& $\displaystyle{\frac{(4r_s - r_c)r_c}{3}}$
& 3.936 & 3.778~\cite{Ebert:2004ck} \\
$\Omega_{bc}$  & $s\{bc\}$ & $1/2^+$
& $\displaystyle{\frac{2 (r_b+r_c) r_s - r_br_c}{3}}$
& 7.257 & 7.088~\cite{Ebert:2004ck} \\
$\Omega'_{bc}$ & $s[bc]$   & $1/2^+$ & $r_br_c$
& 7.296 & 7.116~\cite{Ebert:2004ck} \\
$\Omega_{bb}$  & $s\{bb\}$ & $1/2^+$
& $\displaystyle{\frac{(4r_s - r_b)r_b}{3}}$
& 10.622 & 10.359~\cite{Ebert:2004ck} \\
$\Omega_{cc}^\ast$  & $s\{cc\}$ & $3/2^+$
& $\displaystyle{\frac{(2 r_s + r_c)r_c}{3}}$
& 3.982 & 3.872~\cite{Ebert:2004ck}\\
$\Omega_{bc}^\ast$  & $s\{bc\}$ & $3/2^+$
& $\displaystyle{\frac{r_br_s + r_cr_s + r_br_c}{3}}$
& 7.310 & 7.130~\cite{Ebert:2004ck}\\
$\Omega_{bb}^\ast$  & $s\{bb\}$ & $3/2^+$
& $\displaystyle{\frac{(2 r_s + r_b)r_b}{3}}$
& 10.638 & 10.389~\cite{Ebert:2004ck}\\
$\Omega_{ccb}$ & $b\{cc\}$ & $1/2^+$
& $\displaystyle{\frac{(4r_b - r_c)r_c}{3}}$
& 8.469 & 8.018
\cite{Martynenko:2007je}\\
$\Omega_{cbb}$ & $c\{bb\}$ & $1/2^+$
& $\displaystyle{\frac{(4r_c - r_b)r_b}{3}}$
& 11.801 & 11.280
\cite{Martynenko:2007je}\\
$\Omega_{ccc}^\ast$ & $ccc$ & $3/2^+$ & $r_c^2$
& 5.144 & 4.803
\cite{Martynenko:2007je}\\
$\Omega_{ccb}^\ast$ & $b\{cc\}$ & $3/2^+$
& $\displaystyle{\frac{(2r_b + r_c)r_c}{3}}$
& 8.475 & 8.025
\cite{Martynenko:2007je}\\
$\Omega_{cbb}^\ast$ & $c\{bb\}$ & $3/2^+$
& $\displaystyle{\frac{(2r_c + r_b)r_b}{3}}$
& 11.806  & 11.287
\cite{Martynenko:2007je}\\
$\Omega_{bbb}^\ast$ & $bbb$ & $3/2^+$  & $r_b^2$
& 15.139 & 14.569
\cite{Martynenko:2007je}\\
\hline
\end{tabular}
\end{center}
\end{table}

\newpage

\begin{table}
\begin{center}
\caption{Mass spectrum (in GeV) of $\Lambda_Q$, $\Sigma_Q$, $\Xi_Q$,
$\Xi^\prime_Q$ and $\Omega_Q$ heavy baryon families
with $J^P=\frac{1}{2}^+$}

\vspace*{.25cm}

\def\arraystretch{1.5}
\begin{tabular}{|c|c|c||c|c|c|}
      \hline
      Baryon $(nL)$   & Our results & Ref.~\cite{Ebert:2011kk} &
      Baryon $(nL)$   & Our results & Ref.~\cite{Ebert:2011kk} \\
      \hline
      $\Lambda_c(1S)$ & 2.286       & 2.286 &
      $\Lambda_b(1S)$ & 5.620       & 5.620 \\
      $\Lambda_c(2S)$ & 2.719       & 2.769 &
      $\Lambda_b(2S)$ & 6.053       & 6.089 \\
      $\Lambda_c(3S)$ & 3.045       & 3.130 &
      $\Lambda_b(3S)$ & 6.379       & 6.455 \\
      $\Lambda_c(4S)$ & 3.317       & 3.437 &
      $\Lambda_b(4S)$ & 6.651       & 6.756 \\
      $\Sigma_c(1S)$ & 2.446        & 2.443 &
      $\Sigma_b(1S)$ & 5.809        & 5.808 \\
      $\Sigma_c(2S)$ & 2.833        & 2.901 &
      $\Sigma_b(2S)$ & 6.188        & 6.213 \\
      $\Sigma_c(3S)$ & 3.138        & 3.271 &
      $\Sigma_b(3S)$ & 6.490        & 6.575 \\
      $\Sigma_c(4S)$ & 3.400        & 3.581 &
      $\Sigma_b(4S)$ & 6.748        & 6.869 \\
      $\Xi_c(1S)$ & 2.584       & 2.476 &
      $\Xi_b(1S)$ & 5.940       & 5.803 \\
      $\Xi_c(2S)$ & 2.980       & 2.959 &
      $\Xi_b(2S)$ & 6.331       & 6.266 \\
      $\Xi_c(3S)$ & 3.290       & 3.323 &
      $\Xi_b(3S)$ & 6.638       & 6.601 \\
      $\Xi_c(4S)$ & 3.553       & 3.632 &
      $\Xi_b(4S)$ & 6.900       & 6.913 \\
      $\Xi_c^\prime(1S)$ & 2.613       & 2.579 &
      $\Xi_b^\prime(1S)$ & 5.972       & 5.936 \\
      $\Xi_c^\prime(2S)$ & 3.002       & 2.983 &
      $\Xi_b^\prime(2S)$ & 6.354       & 6.329 \\
      $\Xi_c^\prime(3S)$ & 3.308       & 3.377 &
      $\Xi_b^\prime(3S)$ & 6.657       & 6.687 \\
      $\Xi_c^\prime(4S)$ & 3.569       & 3.695 &
      $\Xi_b^\prime(4S)$ & 6.916       & 6.978 \\
      $\Omega_c(1S)$ & 2.785       & 2.698 &
      $\Omega_b(1S)$ & 6.139       & 6.064 \\
      $\Omega_c(2S)$ & 3.175       & 3.088 &
      $\Omega_b(2S)$ & 6.524       & 6.450 \\
      $\Omega_c(3S)$ & 3.481       & 3.489 &
      $\Omega_b(3S)$ & 6.828       & 6.804 \\
      $\Omega_c(4S)$ & 3.742       & 3.814 &
      $\Omega_b(4S)$ & 7.087       & 7.091 \\
\hline
\end{tabular}
\end{center}

\begin{center}
\caption{Mass spectrum (in GeV) of $\Sigma_Q^\ast$, $\Xi_Q^\ast$,
and $\Omega_Q^\ast$ heavy baryon families
with $J^P=\frac{3}{2}^+$}

\vspace*{.25cm}

\def\arraystretch{1.5}
\begin{tabular}{|c|c|c||c|c|c|}
      \hline
      Baryon $(nL)$   & Our results & Ref.~\cite{Ebert:2011kk} &
      Baryon $(nL)$   & Our results & Ref.~\cite{Ebert:2011kk} \\
      \hline
      $\Sigma_c^\ast(1S)$ & 2.511       & 2.519 &
      $\Sigma_b^\ast(1S)$ & 5.831       & 5.834 \\
      $\Sigma_c^\ast(2S)$ & 2.881       & 2.936 &
      $\Sigma_b^\ast(2S)$ & 6.205       & 6.226 \\
      $\Sigma_c^\ast(3S)$ & 3.178       & 3.293 &
      $\Sigma_b^\ast(3S)$ & 6.504       & 6.583 \\
      $\Sigma_c^\ast(4S)$ & 3.434       & 3.598 &
      $\Sigma_b^\ast(4S)$ & 6.760       & 6.876 \\
      $\Xi_c^\ast(1S)$    & 2.669       & 2.649 &
      $\Xi_b^\ast(1S)$    & 5.991       & 5.963 \\
      $\Xi_c^\ast(2S)$    & 3.043       & 3.026 &
      $\Xi_b^\ast(2S)$    & 6.369       & 6.342 \\
      $\Xi_c^\ast(3S)$    & 3.343       & 3.396 &
      $\Xi_b^\ast(3S)$    & 6.669       & 6.695 \\
      $\Xi_c^\ast(4S)$    & 3.600       & 3.709 &
      $\Xi_b^\ast(4S)$    & 6.927       & 6.984 \\
      $\Omega_c^\ast(1S)$ & 2.831       & 2.768 &
      $\Omega_b^\ast(1S)$ & 6.155       & 6.088 \\
      $\Omega_c^\ast(2S)$ & 3.209       & 3.123 &
      $\Omega_b^\ast(2S)$ & 6.535       & 6.461 \\
      $\Omega_c^\ast(3S)$ & 3.509       & 3.510 &
      $\Omega_b^\ast(3S)$ & 6.837       & 6.811 \\
      $\Omega_c^\ast(4S)$ & 3.767       & 3.830 &
      $\Omega_b^\ast(4S)$ & 7.096       & 7.096 \\
\hline
\end{tabular}
\end{center}
\end{table}


\begin{thebibliography}{99}

\bibitem{Maldacena:1997re}
  J.~M.~Maldacena,
  Adv.\ Theor.\ Math.\ Phys.\  {\bf 2}, 231 (1998)
  [Int.\ J.\ Theor.\ Phys.\  {\bf 38}, 1113 (1999)]
  [arXiv:hep-th/9711200];  
  S.~S.~Gubser, I.~R.~Klebanov and A.~M.~Polyakov,
  Phys.\ Lett.\  B {\bf 428}, 105 (1998)
  [arXiv:hep-th/9802109];
  E.~Witten,
  Adv.\ Theor.\ Math.\ Phys.\  {\bf 2}, 253 (1998)
  [arXiv:hep-th/9802150].
%
%
%
\bibitem{Hard_wall1}
  J.~Polchinski and M.~J.~Strassler,
  Phys.\ Rev.\ Lett.\  {\bf 88}, 031601 (2002)
  [arXiv:hep-th/0109174];
  H.~Boschi-Filho and N.~R.~F.~Braga,
  JHEP {\bf 0305}, 009 (2003)
  [arXiv:hep-th/0212207];
  G.~F.~de Teramond and S.~J.~Brodsky,
  Phys.\ Rev.\ Lett.\  {\bf 94}, 201601 (2005)
  [arXiv:hep-th/0501022];
  Phys.\ Rev.\ Lett.\  {\bf 102}, 081601 (2009)
  [arXiv:0809.4899 [hep-ph]];
  J.~Erlich, E.~Katz, D.~T.~Son and M.~A.~Stephanov,
  Phys.\ Rev.\ Lett.\  {\bf 95}, 261602 (2005)
  [arXiv:hep-ph/0501128];
  L.~Da Rold and A.~Pomarol,
  Nucl.\ Phys.\  B {\bf 721}, 79 (2005)
  [arXiv:hep-ph/0501218];
  K.~Ghoroku, N.~Maru, M.~Tachibana and M.~Yahiro,
  Phys.\ Lett.\  B {\bf 633}, 602 (2006)
  [arXiv:hep-ph/0510334].
\bibitem{Hard_wall2}
  E.~Katz, A.~Lewandowski and M.~D.~Schwartz,
  Phys.\ Rev.\  D {\bf 74}, 086004 (2006)
  [arXiv:hep-ph/0510388].
\bibitem{Hard_wall3}
  H.~R.~Grigoryan and A.~V.~Radyushkin,
  Phys.\ Lett.\  B {\bf 650}, 421 (2007)
  [arXiv:hep-ph/0703069];
  Phys.\ Rev.\  D {\bf 76}, 115007 (2007)
  [arXiv:0709.0500 [hep-ph]].
%
%
%
\bibitem{Soft_wall1}
  A.~Karch, E.~Katz, D.~T.~Son and M.~A.~Stephanov,
  Phys.\ Rev.\  D {\bf 74}, 015005 (2006)
  [arXiv:hep-ph/0602229].
%
%
%
\bibitem{Soft_wall2aa}
  S.~J.~Brodsky, G.~F.~de Teramond,
  Phys.\ Rev.\ Lett.\  {\bf 96}, 201601 (2006)
  [hep-ph/0602252].
\bibitem{Soft_wall2ab}
  S.~J.~Brodsky and G.~F.~de Teramond,
  Phys.\ Rev.\  D {\bf 77}, 056007 (2008)
  [arXiv:0707.3859 [hep-ph]].
\bibitem{Soft_wall2ac}
  S.~J.~Brodsky, G.~F.~de Teramond and A.~Deur,
  Phys.\ Rev.\  D {\bf 81}, 096010 (2010)
  [arXiv:1002.3948 [hep-ph]];
  G.~F.~de Teramond, S.~J.~Brodsky,
  Nucl.\ Phys.\ Proc.\ Suppl.\  {\bf 199}, 89 (2010).
  [arXiv:0909.3900 [hep-ph]].
\bibitem{Soft_wall2b}
  G.~F.~de Teramond and S.~J.~Brodsky,
  AIP Conf.\ Proc.\  {\bf 1296}, 128 (2010)
  [arXiv:1006.2431 [hep-ph]].
%
%
%
\bibitem{Soft_wall3a}
  O.~Andreev,
  Phys.\ Rev.\  D {\bf 73}, 107901 (2006)
  [arXiv:hep-th/0603170];
  O.~Andreev and V.~I.~Zakharov,
  Phys.\ Rev.\  D {\bf 74}, 025023 (2006)
  [arXiv:hep-ph/0604204].
\bibitem{Soft_wall3b}
  H.~Forkel, M.~Beyer and T.~Frederico,
  JHEP {\bf 0707}, 077 (2007)
  [arXiv:0705.1857 [hep-ph]];
  Int.\ J.\ Mod.\ Phys.\  E {\bf 16}, 2794 (2007)
  [arXiv:0705.4115 [hep-ph]];
  W.~de Paula, T.~Frederico, H.~Forkel and M.~Beyer,
  Phys.\ Rev.\  D {\bf 79}, 075019 (2009)
  [arXiv:0806.3830 [hep-ph]].
\bibitem{Soft_wall3c}
  B.~Galow, E.~Megias, J.~Nian and H.~J.~Pirner,
  Nucl.\ Phys.\  B {\bf 834}, 330 (2010)
  [arXiv:0911.0627 [hep-ph]];
  J.~Nian and H.~J.~Pirner,
  Nucl.\ Phys.\  A {\bf 833}, 119 (2010)
  [arXiv:0908.1330 [hep-ph]].
%
%
%
\bibitem{Soft_wall4a}
  H.~R.~Grigoryan and A.~V.~Radyushkin,
  Phys.\ Rev.\  D {\bf 76}, 095007 (2007)
  [arXiv:0706.1543 [hep-ph]].
\bibitem{Soft_wall4b}
  H.~Forkel,
  Phys.\ Rev.\  D {\bf 78}, 025001 (2008)
  [arXiv:0711.1179 [hep-ph]];
  H.~J.~Kwee and R.~F.~Lebed,
  Phys.\ Rev.\  D {\bf 77}, 115007 (2008)
  [arXiv:0712.1811 [hep-ph]].
  P.~Colangelo, F.~De Fazio, F.~Jugeau and S.~Nicotri,
  Phys.\ Lett.\  B {\bf 652}, 73 (2007)
  [arXiv:hep-ph/0703316];
  T.~Gherghetta, J.~I.~Kapusta and T.~M.~Kelley,
  Phys.\ Rev.\  D {\bf 79}, 076003 (2009)
  [arXiv:0902.1998 [hep-ph]];
  Y.~Q.~Sui, Y.~L.~Wu, Z.~F.~Xie and Y.~B.~Yang,
  Phys.\ Rev.\  D {\bf 81}, 014024 (2010)
  [arXiv:0909.3887 [hep-ph]];
  M.~Fujita, K.~Fukushima, T.~Misumi and M.~Murata,
  Phys.\ Rev.\  D {\bf 80}, 035001 (2009)
  [arXiv:0903.2316 [hep-ph]];
  S.~S.~Afonin,
  Int.\ J.\ Mod.\ Phys.\  A {\bf 25}, 5683 (2010).
  [arXiv:1001.3105 [hep-ph]];
  C.~Marquet, C.~Roiesnel and S.~Wallon,
  JHEP {\bf 1004}, 051 (2010)
  [arXiv:1002.0566 [hep-ph]];
  H.~R.~Grigoryan, P.~M.~Hohler and M.~A.~Stephanov,
  Phys.\ Rev.\  D {\bf 82}, 026005 (2010)
  [arXiv:1003.1138 [hep-ph].
%
%
%
\bibitem{Soft_wall5a}
  A.~Vega and I.~Schmidt,
  Phys.\ Rev.\  D {\bf 78}, 017703 (2008)
  [arXiv:0806.2267 [hep-ph]].
\bibitem{Soft_wall5b}
  A.~Vega and I.~Schmidt,
  Phys.\ Rev.\  D {\bf 79}, 055003 (2009)
  [arXiv:0811.4638 [hep-ph]];
  A.~Vega, I.~Schmidt, T.~Branz, T.~Gutsche and V.~E.~Lyubovitskij,
  Phys.\ Rev.\  D {\bf 80}, 055014 (2009)
  [arXiv:0906.1220 [hep-ph]];
  A.~Vega, I.~Schmidt,
  Phys.\ Rev.\  D {\bf 82}, 115023 (2010)
  [arXiv:1005.3000 [hep-ph]].
%
%
%
\bibitem{Soft_wall6}
  S.~J.~Brodsky and G.~F.~de Teramond,
  arXiv:0802.0514 [hep-ph];
  G.~F.~de Teramond and S.~J.~Brodsky,
  AIP Conf.\ Proc.\  {\bf 1257}, 59 (2010)
  [arXiv:1001.5193 [hep-ph]].
%
%
%
\bibitem{Soft_wall7}
  Z.~Abidin and C.~E.~Carlson,
  Phys.\ Rev.\  D {\bf 79}, 115003 (2009)
  [arXiv:0903.4818 [hep-ph]].
%
%
%
\bibitem{Soft_wall8}
  T.~Branz, T.~Gutsche, V.~E.~Lyubovitskij, I.~Schmidt, A.~Vega,
  Phys.\ Rev.\  D {\bf 82}, 074022 (2010)
  [arXiv:1008.0268 [hep-ph]].
%
%
%
\bibitem{Soft_wall9}
  A.~Vega, I.~Schmidt, T.~Gutsche, V.~E.~Lyubovitskij,
  Phys.\ Rev.\  D {\bf 83}, 036001 (2011)
  [arXiv:1010.2815 [hep-ph]].
%
%
%
\bibitem{SWminus}
  F.~Zuo,
  Phys.\ Rev.\  D {\bf 82}, 086011 (2010).
  [arXiv:0909.4240 [hep-ph]];
  S.~Nicotri,
  AIP Conf.\ Proc.\  {\bf 1317}, 322 (2011)
  [arXiv:1009.4829 [hep-ph]].

\bibitem{SWplus}
  A.~Karch, E.~Katz, D.~T.~Son, M.~A.~Stephanov,
  JHEP {\bf 1104}, 066 (2011)
  [arXiv:1012.4813 [hep-ph]].

\bibitem{No_wall}
  S.~S.~Afonin,
  Int.\ J.\ Mod.\ Phys.\  A {\bf 26}, 3615 (2011)
  [arXiv:1012.5065 [hep-ph]].

\bibitem{LFH}
  S.~J.~Brodsky, G.~F.~de Teramond,
  Phys.\ Lett.\  B {\bf 582}, 211 (2004)
  [hep-th/0310227];
  Phys.\ Rev.\  D {\bf 78}, 025032 (2008)
  [arXiv:0804.0452 [hep-ph]].

\bibitem{Randall:1999ee}
  L.~Randall and R.~Sundrum,
  Phys.\ Rev.\ Lett.\  {\bf 83}, 3370 (1999)
  [arXiv:hep-ph/9905221].

\bibitem{Henningson:1998cd}
  M.~Henningson and K.~Sfetsos,
  Phys.\ Lett.\  B {\bf 431}, 63 (1998)
  [arXiv:hep-th/9803251];
  W.~Mueck and K.~S.~Viswanathan,
  Phys.\ Rev.\  D {\bf 58}, 106006 (1998)
  [arXiv:hep-th/9805145];
  R.~Contino and A.~Pomarol,
  JHEP {\bf 0411}, 058 (2004)
  [arXiv:hep-th/0406257].
\bibitem{Buchbinder:2001bs}
  I.~L.~Buchbinder, A.~Pashnev and M.~Tsulaia,
  Phys.\ Lett.\  B {\bf 523}, 338 (2001)
  [arXiv:hep-th/0109067];
  R.~R.~Metsaev,
  Phys.\ Lett.\  B {\bf 590}, 95 (2004)
  [arXiv:hep-th/0312297];
  C.~Germani and A.~Kehagias,
  Nucl.\ Phys.\  B {\bf 725}, 15 (2005)
  [arXiv:hep-th/0411269].
\bibitem{Close:bt}
F.~E.~Close, {\it An Introduction To Quarks And Partons},
(Academic Press, New York, 1979).
%
\bibitem{Donoghue:dd}
J.~F.~Donoghue, E.~Golowich and B.~R.~Holstein,
{\it Dynamics Of The Standard Model},
Cambridge Monogr.\ Part.\ Phys.\ Nucl.\ Phys.\ Cosmol.\
{\bf 2}, 1 (1992).
\bibitem{Inoue:2004jb}
  T.~Inoue, V.~E.~Lyubovitskij, T.~Gutsche and A.~Faessler,
  Int.\ J.\ Mod.\ Phys.\  E {\bf 15}, 121 (2006)
  [arXiv:hep-ph/0404051].
\bibitem{Karliner:2008sv}
  M.~Karliner, B.~Keren-Zur, H.~J.~Lipkin and J.~L.~Rosner,
  Annals Phys.\  {\bf 324}, 2 (2009)
  [arXiv:0804.1575 [hep-ph]].
\bibitem{Faessler:2005gd}
  A.~Faessler, T.~Gutsche, V.~E.~Lyubovitskij and K.~Pumsa-ard,
  Phys.\ Rev.\  D {\bf 73}, 114021 (2006)
  [arXiv:hep-ph/0511319].
\bibitem{PDG}
  K.~Nakamura {\it et al.} (Particle Data Group),
  J.\ Phys.\ G {\bf 37}, 075021 (2010).
\bibitem{Ebert:2004ck}
  D.~Ebert, R.~N.~Faustov, V.~O.~Galkin, A.~P.~Martynenko,
  Phys.\ Rev.\  D {\bf 70}, 014018 (2004) 
  [hep-ph/0404280].
\bibitem{Ebert:2011kk}
  D.~Ebert, R.~N.~Faustov, V.~O.~Galkin,
  Phys.\ Rev.\  D {\bf 84}, 014025 (2011) 
  [arXiv:1105.0583 [hep-ph]].
\bibitem{Martynenko:2007je}
  A.~P.~Martynenko,
  Phys.\ Lett.\  B {\bf 663}, 317-321 (2008) 
  [arXiv:0708.2033 [hep-ph]].
\bibitem{LCQM1}
  M.~A.~Ivanov, V.~E.~Lyubovitskij, J.~G.~K\"orner and P.~Kroll,
  Phys.\ Rev.\ D {\bf 56}, 348 (1997)
  [arXiv:hep-ph/9612463];
  M.~A.~Ivanov, J.~G.~K\"orner, V.~E.~Lyubovitskij and A.~G.~Rusetsky,
  Phys.\ Rev.\  D {\bf 57}, 5632 (1998)
  [arXiv:hep-ph/9709372];
  M.~A.~Ivanov, J.~G.~K\"orner, V.~E.~Lyubovitskij and A.~G.~Rusetsky,
  Phys.\ Rev.\ D {\bf 60}, 094002 (1999)
  [arXiv:hep-ph/9904421];
  A.~Faessler, T.~Gutsche, M.~A.~Ivanov, J.~G.~K\"orner,
  V.~E.~Lyubovitskij, D.~Nicmorus and K.~Pumsa-ard,
  Phys.\ Rev.\ D {\bf 73}, 094013 (2006)
  [arXiv:hep-ph/0602193];
  A.~Faessler, T.~Gutsche, M.~A.~Ivanov, J.~G.~Korner, V.~E.~Lyubovitskij,
  Phys.\ Rev.\  D {\bf 80}, 034025 (2009)
  [arXiv:0907.0563 [hep-ph]];
  T.~Branz, A.~Faessler, T.~Gutsche, M.~A.~Ivanov,
  J.~G.~Korner, V.~E.~Lyubovitskij, B.~Oexl,
  Phys.\ Rev.\  D {\bf 81}, 114036 (2010)
  [arXiv:1005.1850 [hep-ph]].
\bibitem{LCQM2}
 M.~A.~Ivanov, M.~P.~Locher and V.~E.~Lyubovitskij,
  Few Body Syst.\  {\bf 21}, 131 (1996);
  A.~Faessler, T.~Gutsche, B.~R.~Holstein, V.~E.~Lyubovitskij,
  D.~Nicmorus and K.~Pumsa-ard,
  Phys.\ Rev.\ D {\bf 74}, 074010 (2006)
  [arXiv:hep-ph/0608015];
  A.~Faessler, T.~Gutsche, B.~R.~Holstein, M.~A.~Ivanov,
  J.~G.~K\"orner and V.~E.~Lyubovitskij,
  Phys.\ Rev.\  D {\bf 78}, 094005 (2008)
  [arXiv:0809.4159 [hep-ph]]. 

\end{thebibliography}
\end{document}